\documentclass[12pt,aps,prd,a4paper,superscriptaddress,nofootinbib,onecolumn]{revtex4-2}
\usepackage{amsfonts}
\usepackage{amsmath}
\usepackage{babel}
\usepackage[active]{srcltx}
\usepackage{graphicx}
\usepackage{color}
\usepackage{float}
\usepackage{changebar}
\usepackage{esint}
\usepackage{dsfont}
\usepackage{ae}
\usepackage{multirow}
\usepackage[usenames,dvipsnames]{xcolor}
\usepackage[colorlinks=true,pdfstartview=FitV,linkcolor=blue,citecolor=blue,urlcolor=blue,breaklinks=true]{hyperref}
\usepackage{braket}
\usepackage{mathtools}
\usepackage{slashed}
\usepackage{ulem}
\usepackage{empheq}
\usepackage{multirow}
\usepackage[utf8]{inputenc}
\usepackage[T1]{fontenc}

\makeatletter
\RequirePackage{color}

\begin{document}

\title{Physical aspects of the deformation of $\mathbb{Z}_2$ kinks in a generalized $\phi^4$ model}
\author{F. C. E. Lima}
\email{cleiton.estevao@fisica.ufc.br}
\email{fcelima.fisica@gmail.com}
\affiliation{Departamento de F\'{\i}sica, Universidade Federal do Maranh\~{a}o, Campus
Universit\'{a}rio do Bacanga, S\~{a}o Lu\'{\i}s (MA), 65080-805, Brazil.}
\author{C. A. S. Almeida}
\email{carlos@fisica.ufc.br}
\email{carllin.almeida@gmail.com}
\affiliation{Departamento do F\'{\i}sica, Universidade Federal do Cear\'{a}, 60455-760,
Fortaleza, CE, Brazil.}
\affiliation{Institute of Cosmology, Department of Physics and Astronomy, Tufts
University, Medford, Massachusetts 02155, USA.}
\author{Rodolfo Casana}
\email{rodolfo.casana@ufma.br}\email{rodolfo.casana@gmail.com}
\affiliation{Departamento de F\'{\i}sica, Universidade Federal do Maranh\~{a}o, Campus
Universit\'{a}rio do Bacanga, S\~{a}o Lu\'{\i}s (MA), 65080-805, Brazil.}

\begin{abstract}
\vspace{0.5cm}
We study a generalized $\phi^4$ model that gives rise to BPS kink/antikink configurations with compacton-like profiles. One observes that the positive parameter controlling the generalizing function promotes an infinity degenerescence of the BPS solutions. We then use the Differential Configurational Complexity technique to distinguish the degenerate configurations, which allows us to obtain the parameter values providing the most likely field profiles. Besides, the analysis of the excitation spectrum of the model shows the existence of translational and vibrational modes. Thus, the emergence of bound states of solitons (bions) and resonance phenomena is guaranteed when analyzing the scattering of kink/antikink structures. In this way, one notes that depending on the initial velocity, the collision can be inelastic or quasi-elastic, even in the case of compacton-like configurations.

\end{abstract}
\maketitle

\thispagestyle{empty}

\newpage
\section{Introduction}

Since 1966, through the significant work of Finkelstein \cite{Finkelstein}, it has been known that in (1+1)-dimensional nonlinear field theories give rise to topological extended objects called kinks. Furthermore, topological structures emerging in some (4,4)-supersymmetric (1+1)-dimensional sigma models are called Q-kinks \cite{Abraham}. Adjacent to these studies, the kinks arise in several areas of knowledge. For instance, one can find research on the regularization from the topological sectors of these structures in Ref. \cite{Evslin}. Meanwhile, in Ref. \cite{Alonso}, the interaction between the shape modes of oscillating kinks arising in scalar field theory models with two components
is studied. These structures also emerge in condensed matter physics, where one examines the electronic properties of Van der Waals insulators \cite{Lee}. Conscious of the multiple applications of these structures, our purpose is to investigate them by assuming a theory with noncanonical dynamics. Thus, two questions naturally arise: Would it be possible to obtain kinks in a noncanonical theory, and what influence do noncanonical dynamics have on these structures?

After the discovery of solitary waves in 1834 \cite{Vachaspati}, announced by John Scott Russell, several studies emerged aiming to elucidate novel
characteristics of these configurations. Particularly noteworthy is that they also present structures with finite wavelengths called compactons \cite{Rosenau}. Concurrently, it is possible to transform a kink into a compacton-like profile by using some adequate mechanism. In this sense, the Ref. \cite{Dusuel} stands as a pioneer in studies about deformation mechanisms of $\phi^4$-kinks into compacton-like profiles. Subsequently, in the static case, employing a deformable potential approach, one can transform kink structures into compacton-like configurations \cite{Bazeia}.

Among the methods responsible for the contraction or deformation of kinks, one considers the generalized theories and the $k$-generalized or noncanonical models. In this context, the first ones have a kinetic term coupled to a generalized function associated with the symmetry-breaking potential \cite{CSantos1, CSantos2, CAdam}. Predominantly, the generalized solutions manifest behaviors analogous to the structures similar to conventional theory. However, in the $k$-generalized models, the exotic dynamics introduce variations in the aspects of these structures, notably modifying their amplitudes \cite{Babichev1, Andrews,LimaAlmeida}. Thus, these theories are relevant once they allow the appearance of a diversity of new characteristics for the solutions. For instance, they can provide explanations concerning the accelerated inflationary phase of the universe \cite{Picon1}, the description of gravitational waves \cite{Mukhanov}, and the nature of dark matter \cite{Picon2}.

A second analysis concerns employing arguments derived from Configurational Entropy (CE) \cite{Gleiser1, Gleiser2} that allow us to distinguish and delineate the most likely structures that may emerge in the model. Specifically, to reach our objective, one uses a variant of the CE called Differential Configurational Complexity (DCC) \cite{Gleiser5} that quantifies the complexity underlying the construction of a configuration in a field model \cite{Gleiser3, Gleiser4, Gleiser6, LimaAlmeida}.

We finalized our study by discussing the scattering process of compacton-like structures. Since the 1970s \cite{Belendryasova, Belova}, the phenomenon of kink scattering and similar structures has attracted enough attention from the academic community. Among the techniques developed to analyze this phenomenon, we mention the analytical collective coordinate approach  \cite{Gani1, Christov, Takyi, Demirkaya}, and also the numerical methods have recently become a powerful tool for studying the dynamics of such phenomenon \cite{Dorey, Marjaneh, Gani2}. For instance, resonance phenomena, such as escape windows and quasi-resonances, have been discovered through numerical methods of kink dispersion. In this context, Ref. \cite{Dusuel} shows the compacton-like profiles' collision process but does not report the appearance of configurations called bions\footnote{The bions \cite{Anninos} are ``temporary bound-states" formed during the kink-antikink collision process, and they possess a topological charge, a combination of the topological charges of the original configurations.}. Thus, we will seek structures emergent during the scattering process of the generalized kinks found by us.

The manuscript is delineated as follows: Section \ref{sec2} studies the BPS structure of a (1+1)-dimensional generalized scalar field model and the emerging structures endowed with BPS properties. The principal aim of our inquiry is to analyze the transition of the kink solutions towards a compacton-like profile. In Sec. \ref{sec4}, one presents a study on the DCC, seeking to discern the most likely BPS field configurations inherent to the proposed theory. In Sec. \ref{sec3}, one analyzes both the excitation spectrum and the scattering phenomena of the solutions. Finally, we present our remarks and conclusions in Sec. \ref{sec5}.

\section{Generalized scalar model: BPS formalism \label{sec2}}

We will focus our efforts on constructing a generalized scalar field theory supporting a BPS structure. Thus, let us start our work by considering the following action in two-dimensional flat spacetime\footnote{Throughout the paper, the metric adopted is $\eta_{\mu \nu}=\text{diag}(1,-1)$ and the Greek indexes run on $0$ or $1$.}, i.e.,
\begin{equation}
\mathcal{S}=\int \,d^{2}x\,\left[ \frac{f(\phi )}{2}\partial _{\mu }\phi
\partial ^{\mu }\phi -V(\phi )\right] .  \label{Saction}
\end{equation}%
Here, $\phi \equiv \phi \left( x\right) $ is a scalar field, $f(\phi )$ is a nonnegative generalizing function, and $V(\phi )$ represents the self-interacting potential, which will be determined during the implementation of the BPS formalism.

The equation of motion concerned with action (\ref{Saction}) is
\begin{equation}
f\partial _{\mu }\partial ^{\mu }\phi +\frac{f_{\phi }}{2}(\partial _{\mu
}\phi )\,\partial ^{\mu }\phi +V_{\phi }=0,  \label{EOM111}
\end{equation}%
concurrently, $f_{\phi }=\frac{\partial f}{\partial \phi }$ and $V_{\phi }=%
\frac{\partial V}{\partial \phi }$.

Considering the static case, the equation of motion assumes the following form:
\begin{equation}
f\frac{d^{2}\phi }{dx^{2}}+\frac{1}{2}f_{\phi }\left( \frac{d\phi }{dx}%
\right) ^{2}=V_{\phi },  \label{EqOfMotion}
\end{equation}%
where now $x$ represents only the position coordinate. In this regime, the total energy reads
\begin{equation}
\mathrm{E}=\int_{-\infty }^{\infty }\,dx\,\left[ \frac{1}{2}f\left( \frac{%
d\phi }{dx}\right) ^{2}+V\right] .  \label{eq4x1}
\end{equation}%
Now, let us analyze whether the model supports BPS property. To reach our purpose, we will introduce auxiliary function $W\equiv W(\phi )$, which plays a relevant role once it is related to the potential and total energy of the model, proving to be an advantageous approach \cite{Vachaspati, Lima4}. Then, the implementation of the BPS formalism allows us to write the total energy as
\begin{equation}
\mathrm{E}=\int_{-\infty }^{\infty }\,dx\,\left[ \frac{1}{2f}\left( f\frac{%
d\phi }{dx}\mp W_{\phi }\right) ^{2}+V-\frac{W_{\phi }^{2}}{2f}\pm \frac{dW}{%
dx}\right] ,
\end{equation}%
where we have defined $W_{\phi }=\frac{\partial W}{\partial \phi }$. Here, we consider that the model will admit a BPS structure whether one assumes
\begin{equation}
V\left( \phi \right) =\frac{W_{\phi }^{2}}{2f(\phi )}.  \label{potx2}
\end{equation}
Consequently, the total energy becomes write as
\begin{equation}
\mathrm{E}=\mathrm{E}_{\text{BPS}}+\int_{-\infty }^{\infty }\,dx\frac{1}{2f%
}\left( f\frac{d\phi }{dx}\mp W_{\phi }\right) ^{2},  \label{energy2}
\end{equation}%
with $\mathrm{E}_{\text{BPS}}$ defining the Bogomol'nyi bound,
\begin{equation}
\mathrm{E}_{\text{BPS}}=\pm \int_{-\infty }^{\infty }\,\frac{dW}{dx}\,dx=\pm %
\left[ W\left( \phi _{+\infty }\right) -W\left( \phi _{-\infty }\right) %
\right] >0.  \label{EB}
\end{equation}
Note that energy has a lower bounded, i.e., $\mathrm{E}\geq \mathrm{E}_{%
\text{BPS}}$, which is saturated ($\mathrm{E}=\mathrm{E}_{\text{BPS}}$) when
the scalar field satisfies\ the differential equation
\begin{equation}
\frac{d\phi }{dx}=\pm \frac{W_{\phi }}{f(\phi )}.  \label{EqBPS}
\end{equation}
The expression (\ref{EqBPS}) establishes the BPS or self-dual equation of the model.

Considering Eq. (\ref{EB}) or Eq. (\ref{eq4x1}) we infer that the BPS energy density is given by
\begin{equation}
\mathcal{E}_{\text{BPS}}=\pm \frac{dW}{dx}=\frac{W_{\phi }^{2}}{f(\phi) },
\label{Ebpsx1}
\end{equation}
where we use the BPS equation to obtain the last expression.

\subsection{Modifying the $\protect\phi^4$ model}

To continue our study, we need to choose both the superpotential $W(\phi)$ and the generalizing function $f(\phi)$ to define a BPS potential (\ref{potx2}) that preserves the $\mathds{Z}_{2}$\footnote{A system possesses a $\mathds{Z}_{2}$ symmetry when the action is invariant under the transformation $\phi\rightarrow-\phi$.} symmetry \cite{Vachaspati,Rajaraman} but at the same time able to promote the spontaneous symmetry breaking of it. This way, for our purpose, we use a $\phi^4$ model \cite{Pereira1,Pereira2} defined by the following superpotential,
\begin{equation}
W(\phi )=\sqrt{\frac{\lambda }{2}}\left(\nu^{2}\phi -\frac{1}{3}\phi
^{3}\right) ,  \label{WPot}
\end{equation}
and we also fix the generalizing function as
\begin{equation}
f(\phi )=\cos \left( m\pi \phi \right) .  \label{fgg}
\end{equation}
The generalizing function preserves the vacuum structure of the $\phi^4$-model. Nevertheless, despite possessing the same Bogomol'ny bound, the BPS solutions are affected by the $m$ parameter that changes the profiles, leading to more localized solutions. Thus, the chosen function allows us to obtain more geometrically constrained topological structures and, consequently, enables the emergence of compacton-like configurations.

\begin{figure}[!ht]
\centering\includegraphics[height=6.5cm,width=7.5cm]{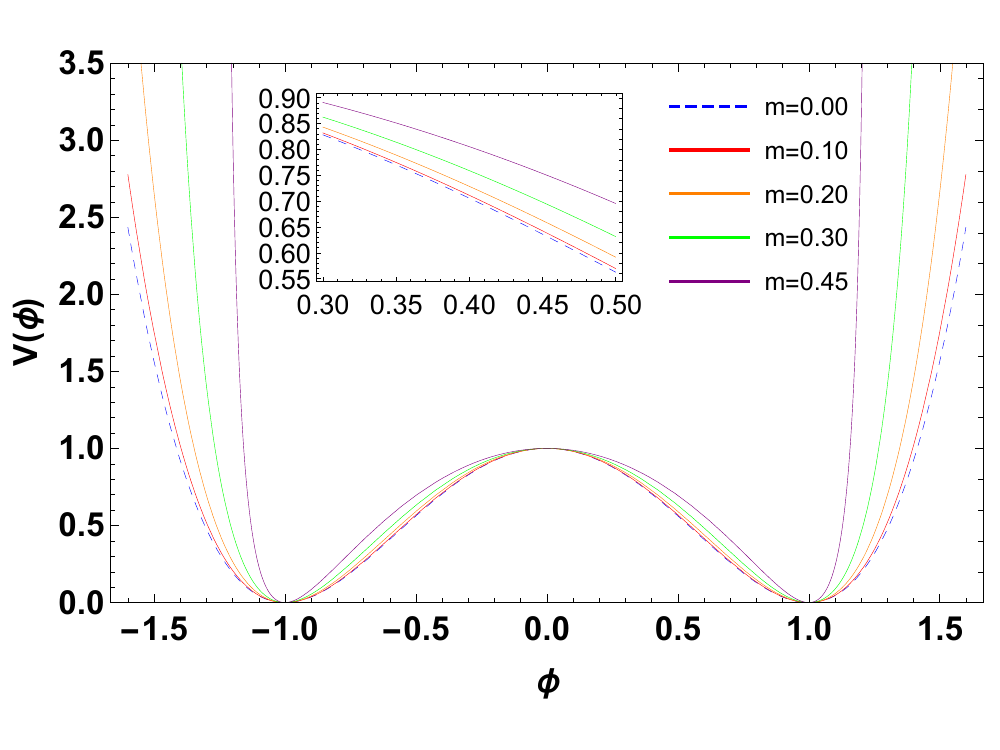} %
\includegraphics[height=6.5cm,width=7.5cm]{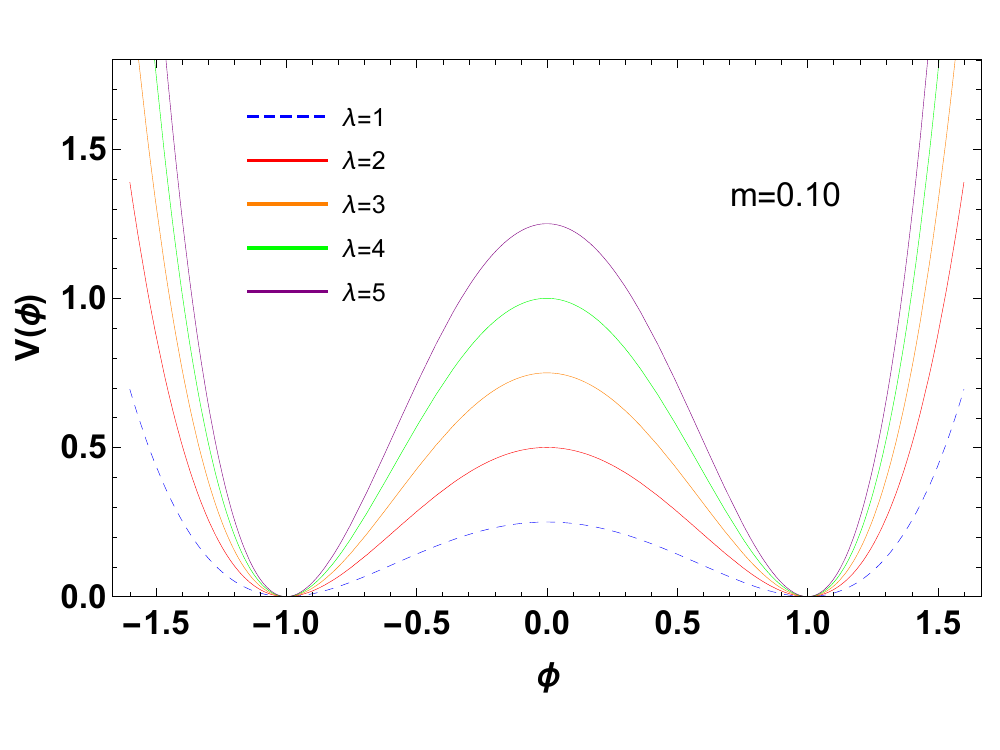} \vspace{-0.7cm}

\hspace{0.65cm} (a) \hspace{6.9cm} (b) \vspace{-0.3cm}
\caption{Potential $V(\phi)$ vs. $\phi$ [Eq. (\ref{PotMod})]: (a) for $\lambda=4$ and $m$ varying. Here, $m=0$ (blue dashed line) represents the standard $\phi^4$-potential. (b) For $m=0.1$ and $\lambda$ varying.} \label{fig1}
\end{figure}

Adopting the expressions (\ref{WPot}) and (\ref{fgg}), one obtains the potential
\begin{equation}
V\left( \phi \right) =\frac{\lambda }{4}\frac{(\nu^{2}-\phi ^{2})^{2}}{\cos
\left( m\pi \phi \right) },  \label{PotMod}
\end{equation}
[whose behavior is shown in Figs. \ref{fig1}(a) and \ref{fig1}(b)] where we will consider $0\leq m<0.5$ and $\nu=1$, $\phi \in \lbrack -1,1]$. We impose these restrictions to ensure that there are no divergences in the potential, thus, preserving the emergence of topological structures. Consequently, one can use these structures to describe specific physical phenomena, such as Josephson junctions \cite{Scott} and systems with one-dimensional dislocations \cite{Lamb}. Furthermore, it is possible to employ this theory in a more theoretical context, allowing for the description of spacetimes with constant negative curvature \cite{Eisenhart} or simply regarding them as objects of intrinsic interest in integrable field theories \cite{Babelon}. 

Adopting the superpotential (\ref{WPot}) and Eq. (\ref{fgg}), the BPS
equation (\ref{EqBPS}) ruling the self-dual solutions becomes
\begin{equation}
\frac{d\phi }{dx}=\pm \sqrt{\frac{\lambda }{2}}\frac{(\nu ^{2}-\phi ^{2})}{%
\cos (m\pi \phi )}.  \label{BPSEq111}
\end{equation}%
Thus, the boundary conditions satisfied by the scalar field to obtain kink and antikink configurations are
\begin{eqnarray}
\lim_{x\rightarrow \pm \infty }\phi (x)=\phi _{\pm \infty }=\pm \nu ,\\[0.2cm]
\lim_{x\rightarrow \mp \infty }\phi (x)=\phi _{\mp \infty }=\mp \nu ,
\end{eqnarray}
respectively, where $\nu >0$, which represent the vacuum value of  $|\phi|$. Thus, one obtains from Eqs. (\ref{EB}) and (\ref{WPot}) the BPS energy as
\begin{equation}
\mathrm{E}_{\text{BPS}}= \frac{2\sqrt{2\lambda }\nu^{3}}{3}. \label{ENbpsx}
\end{equation}

\begin{figure}[!ht]
\centering
\includegraphics[height=6.5cm,width=7.5cm]{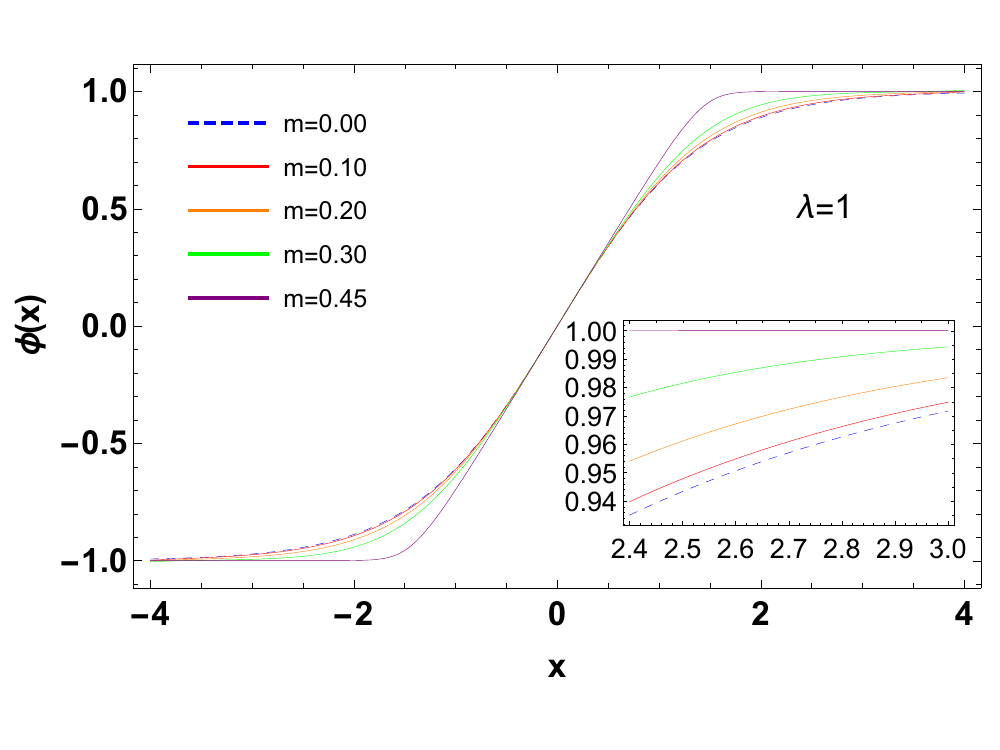} %
\includegraphics[height=6.5cm,width=7.5cm]{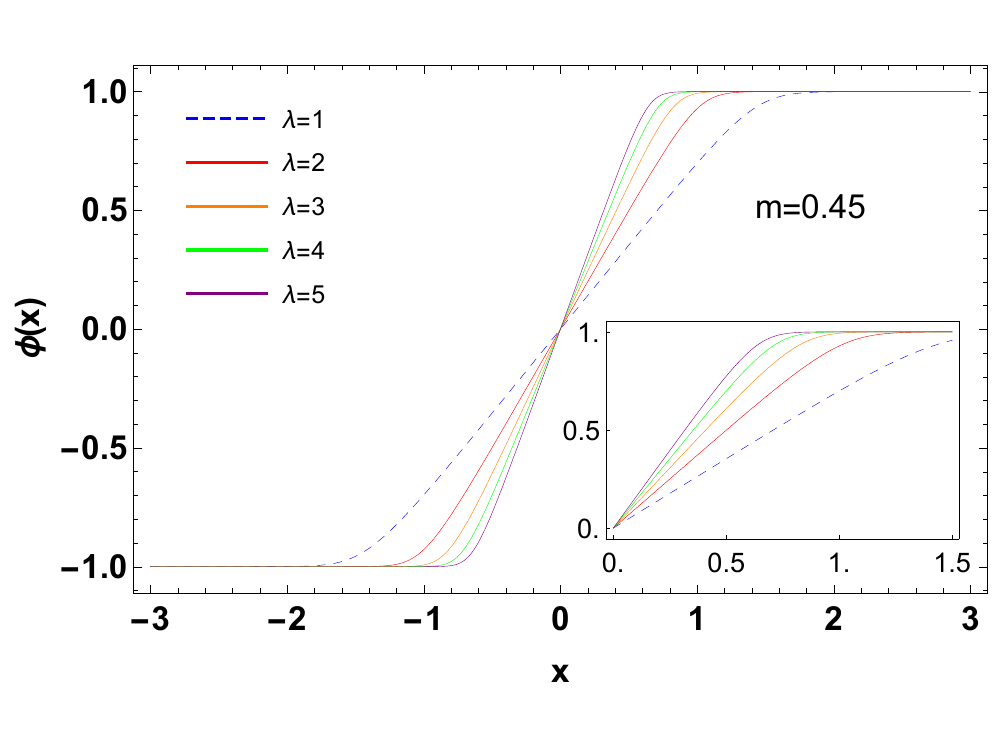}\vspace{-0.7cm}

\hspace{0.85cm} (a) \hspace{6.9cm} (b) \vspace{-0.3cm}
\caption{Kink solution from Eq. (\ref{BPSEq111}): (a) For $\lambda=1$ and $m$ varying. Here, $m=0$ (blue dashed line) represents the kink of the standard $\phi^4$-potential. (b) For $m=0.45$ and $\lambda$ varying. For all the cases, one assumes the VEV is $\nu=1$.} \label{fig2}
\end{figure}

\begin{figure}[!ht]
\centering
\includegraphics[height=6.5cm,width=7.5cm]{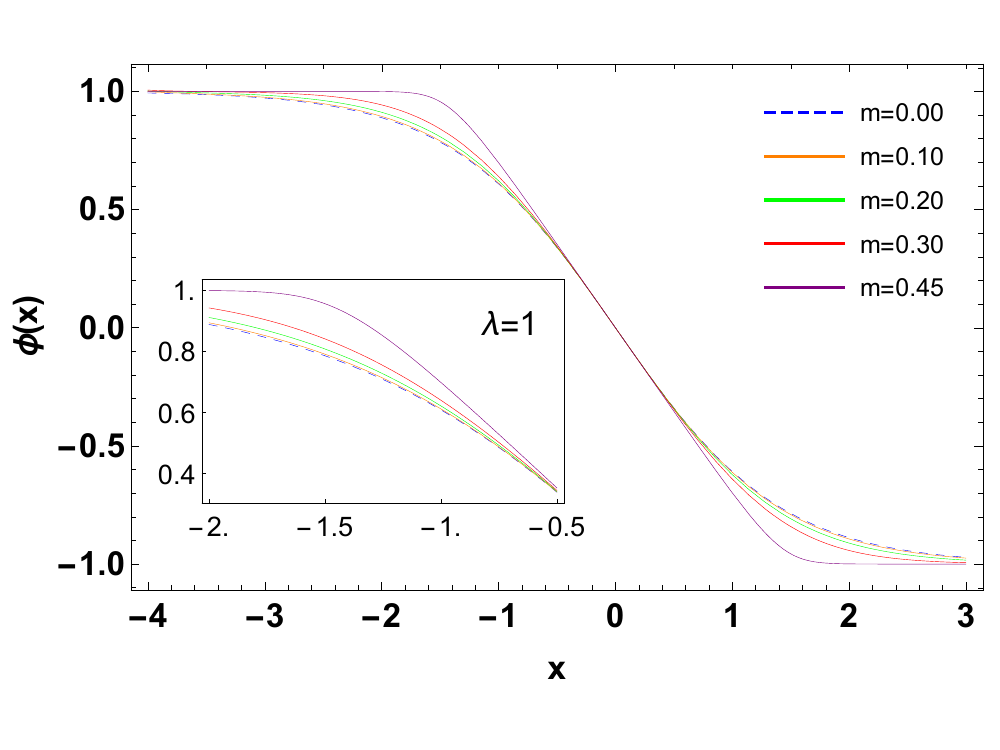} \includegraphics[height=6.5cm,width=7.5cm]{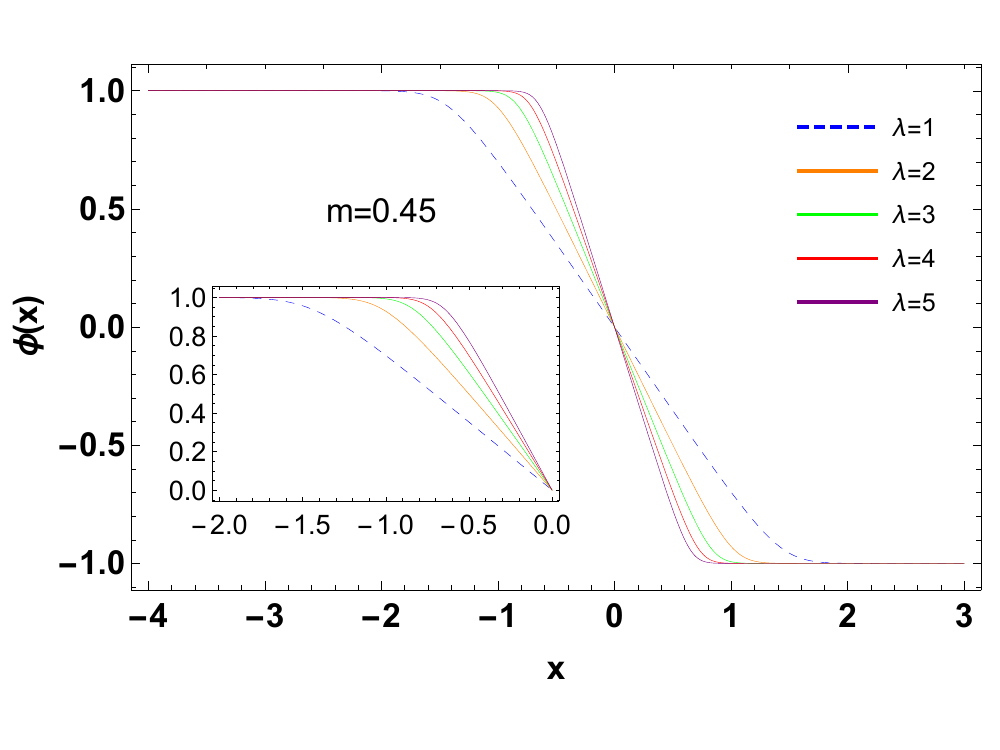}\vspace{-0.7cm}

\hspace{0.85cm} (a) \hspace{6.9cm} (b) \vspace{-0.3cm}
\caption{Antikink solution from Eq. (\ref{BPSEq111}): (a) For $\lambda=1$ and $m$ varying. Here, $m=0$ (blue dashed line) represents the antikink of the standard
$\phi^4$-potential. (b) For $m=0.45$ and $\lambda$ varying. For all the cases, one assumes the VEV is $\nu=1$.} \label{fig3}
\end{figure}

Note that the equation (\ref{BPSEq111}) has the form
\begin{align}
\frac{d\phi(x)}{dx}=g(\phi(x)),
\end{align}
thus, one can discretize the domain of the $x$ independent variable into $n$ points, namely, $x_0, x_1, x_2, \ldots, x_n$; subsequently, we use the interpolation method to estimate the solution $\phi(x)$ at the intermediate points \cite{Hildebrand}. Applying this approach, the numerical solutions of the equation (\ref{BPSEq111}) are calculated, see Figs. \ref{fig2}[(a) and (b)] and \ref{fig3}[(a) and (b)] for kink and antikink configurations, respectively.

\begin{figure}[!ht]
\centering
\includegraphics[height=6.5cm,width=7.5cm]{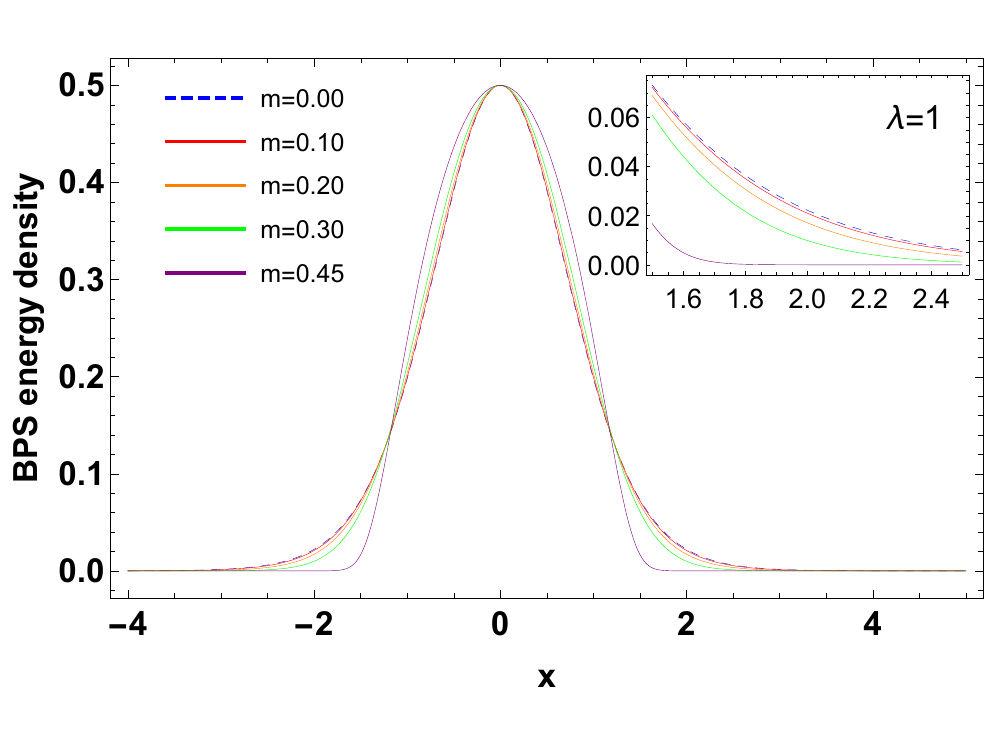}%
\includegraphics[height=6.5cm,width=7.5cm]{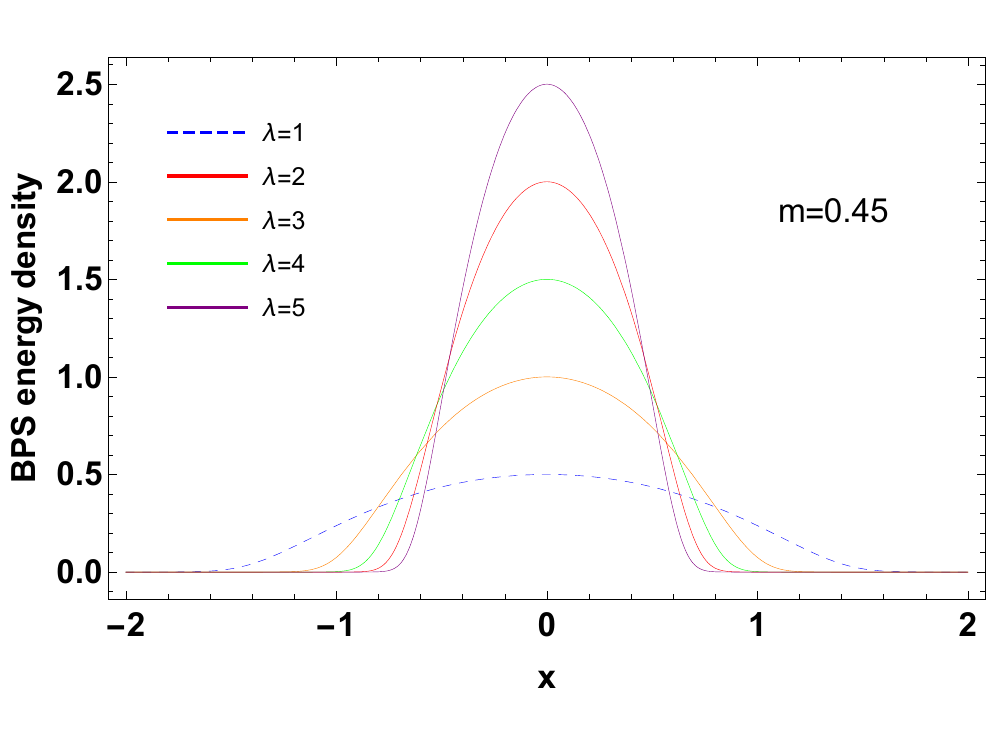}\vspace{-0.6cm}
\begin{center}
\hspace{0.7cm} (a) \hspace{6.9cm} (b) \vspace{-0.3cm} \end{center}
\vspace{-0.6cm}
\caption{The BPS energy density for the kink  and antikink configurations exposed in Figs. \ref{fig2}(a), \ref{fig2}(b), \ref{fig3}(a), and \ref{fig3}(b). In Fig. (a), $m=0$ (blue dashed line) represents the BPS energy density of the kink e anti-kink solutions of the standard $\phi^4$-potential.}
\label{fig4}
\end{figure}

The energy density (\ref{Ebpsx1}) results in
\begin{equation}
\mathcal{E}_{\text{BPS}}(x)=\frac{\lambda }{2}\frac{(\nu^2-\phi^{2})^{2}}{
\cos\left( m\pi \phi \right)}.  \label{EB1}
\end{equation}%
whose numerical profiles are depicted in Fig. \ref{fig4} for kink and antikink configurations.

The results displayed in figures \ref{fig2}, \ref{fig3}, and \ref{fig4} demonstrate how the kink-like configurations deform to compacton-like structures when the parameter $m$ approaches the value $0.5$. Studying compact configurations, or compactons, is of interest due to their stability. Generally speaking, the compacton stability is similar to one for kinks, and its stability potential is comparable to the P\"oschl-Teller-like.

Therefore, we observe structures with a compact-like profile emerging in the generalized theory when $m\rightarrow 0.5$. It is pertinent to note that the occurrence of these structures is feasible due to the generalization adopted, which modifies the $\phi^4$ symmetry-breaking potential ($m=0$)  according to Eq. (\ref{PotMod}). Thereby, the control of the kink- and antikink-like profiles occurs because the modified potential makes the solutions more massive when $m\rightarrow 0.5$.


\section{The Differential Configurational Complexity (DCC)\label{sec4}}

This section will study through the concepts originating from DCC formalism what is the best set of the parameters $\lambda$ and  $m$ to the arising of the topological kinks engendered from the generalized $\phi^4$ model with the generalizing function $f(\phi)= \cos(m\pi\phi)$. As it is known, in principle, the parameters $m$ and   $\lambda$ can assume any values; nevertheless, the DCC analysis allows one to estimate the most likely values for these parameters. That happens because a lower complexity entropy implies a higher probability of the occurrence of the topological structures \cite{Gleiser5}. One has employed this formalism in investigations involving several systems with localized energy configurations \cite{Gleiser6, Braga1, Braga2, Braga3, Roldao1, Roldao2}. Additionally, it is noteworthy that DCC provides a good tool for studying the mass spectrum of the kaon vector resonances \cite{Roldao3}.

The DCC technique \cite{Gleiser5} presupposes the existence of a modal fraction, with normalization taking into account the square of the contribution of the maximum mode of the wave to the BPS energy density \cite{Bogomol,PrasadS}. Then, to perform the DCC calculation, one must compute the Fourier transform:
\begin{equation}
\mathcal{E}_{\text{BPS}}(\mathbf{k})=\frac{\lambda }{\sqrt{8\pi }}%
\int_{-\infty }^{\infty }\,\bigg\{\frac{[\nu ^{2}-\phi (x)^{2}]^{2}}{\cos
[m\pi \phi (x)]}\bigg\}\,\text{e}^{-i\mathbf{k}\cdot x}\,dx.  \label{EFT}
\end{equation}

Substituting the BPS energy density in the power spectrum (\ref{EFT}), we build the modal fraction at the reciprocal space, i.e.,
\begin{equation}
g(\mathbf{k})=\frac{|\mathcal{E}_{\text{BPS}}(\mathbf{k})|^{2}}{|\mathcal{E}%
_{\text{BPS}}^{\text{(max)}}(\mathbf{k})|^{2}}.  \label{MF}
\end{equation}%
In this scenario, $\mathcal{E}_{\text{BPS}}^{\text{(max)}}(\mathbf{k})$ denotes the BPS energy density at the reciprocal space for the maximum wave mode. Additionally, $g(\mathbf{k})$ represents the modal fraction\footnote{The modal fraction describes the contribution of each wave mode to a power spectral density.} of the model. This modal fraction is minimal for spectral densities of localized structures, informing us about the least complex and most likely configuration. In this case, the DCC is
\begin{equation}
\text{DCC}(\mathbf{k})=-\int_{-\infty }^{\infty }g(\mathbf{k})\,\text{ln}[g(%
\mathbf{k})]\,d\mathbf{k}.  \label{CDCC}
\end{equation}

Now, let us consider the numerical solutions depicted in Figs. \ref{fig2} and \ref{fig3}, and employing a numerical approach\footnote{The numerical method applied was the interpolation method. This method is used with steps of $10^{-4}$ for the range $[-1000,1000]$.}, we compute the integrals (\ref{EFT}) and (\ref{CDCC}). We expose the numerical solutions obtained in Figs. \ref{fig12}(a) and \ref{fig12}(b).

\begin{figure}[!ht]
\centering
\includegraphics[height=4.85cm]{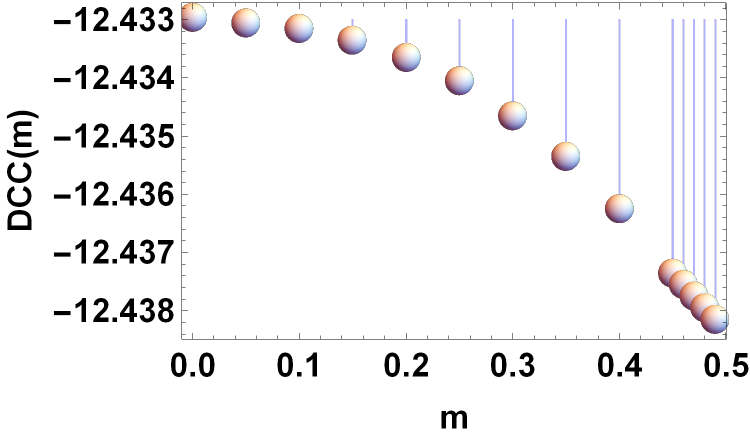}%
\includegraphics[height=4.85cm]{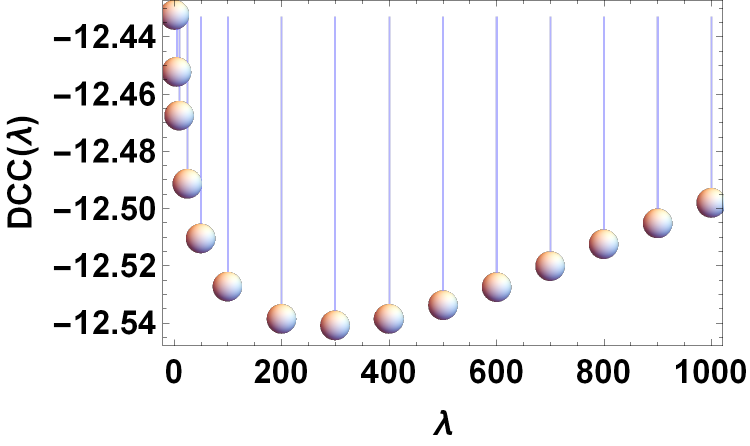}\vspace{-0.3cm}

\hspace{1.6cm} (a) \hspace{7.35cm} (b) \vspace{-0.3cm}
\caption{Numerical result from the DCC. (a) DCC in terms of
the $m$ parameter. (b) DCC in terms of the $\lambda$ parameter.} \label%
{fig12}
\end{figure}

\begin{figure}[!ht]
\centering
\includegraphics[height=6.5cm,width=7cm]{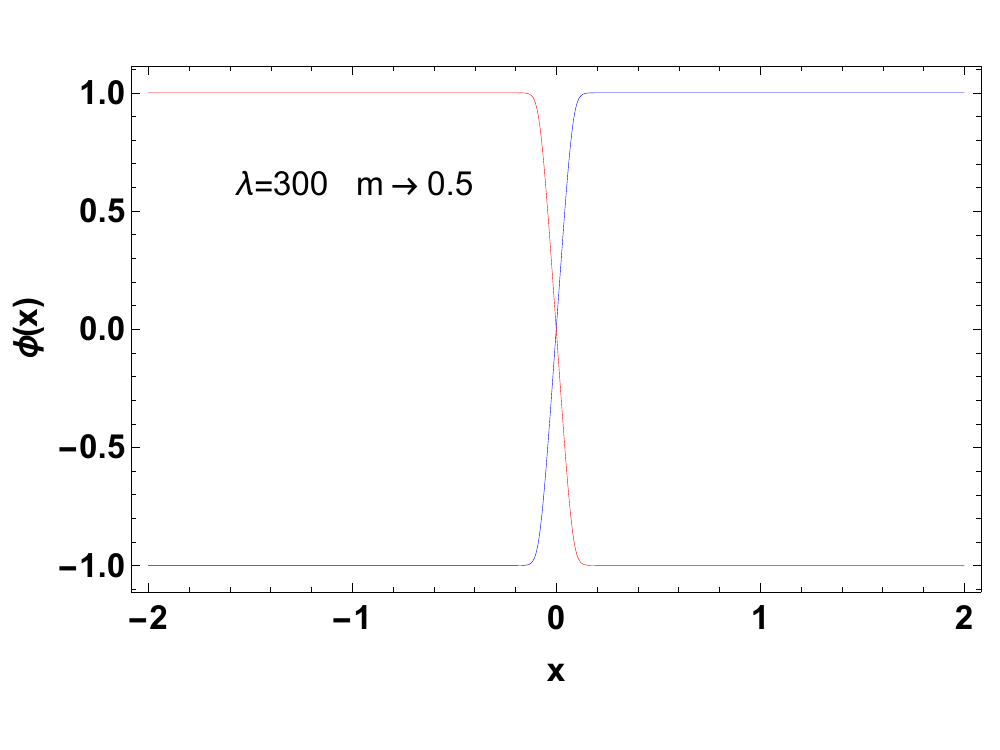} \vspace{-1cm} %
\caption{Field configurations with lower complexity in agreement with DCC.}
\label{fig13}
\end{figure}

The numerical simulations indicate that field configurations with lower complexity, and hence, higher probability, arise when $\lambda \simeq 300$ and $m \to 0.5$. We display the most likely field configurations in Fig. \ref{fig13}. It is noteworthy that configurations with lower complexity, and hence, more likely, correspond to compacton-like configurations, as reported previously in Sec. \ref{sec2}.

\section{Excitation spectra and scattering of structures \label{sec3}}

In this section, we will study the stability of the BPS solutions (i.e., the excitation spectrum) and the scattering of these configurations with opposing topological charges. Our endeavors are motivated by confirming how the generalizing function affects or modifies both the excitation spectrum's nature and the kink-antikink collision process when the structures evolve toward compact-like configurations.

\subsection{Excitation spectra}

To initiate our research, let us begin by studying the stability of the BPS solution from the equation (\ref{EOM111}). To calculate the excitation spectrum, we introduce small perturbations $\delta \phi (x,t)$, i.e.,
\begin{equation}
\phi (x,t)=\phi _{0}(x)+\delta \phi (x,t),\quad \text{with}\quad ||\delta
\phi ||\ll ||\phi _{0}||.  \label{pert}
\end{equation}%
where $\phi _{0}\left( x\right) $ is the kink configuration solving the BPS equation (\ref{EqBPS}). Besides, we consider the perturbation as
\begin{equation}
\delta \phi (x,t)=\eta (x)\cos (\omega t),\quad ||\eta ||\ll ||\phi _{0}||
\end{equation}%
which after substituting into the equation of motion (\ref{EOM111}) and considering only linear contributions in $\eta (x)$, one obtains the following eigenvalue equation
\begin{equation}
-\frac{d^{2}\eta }{dx^{2}}-\left. \frac{1}{f}\frac{\partial f}{\partial \phi
}\right\vert _{\phi =\phi _{0}}\left( \frac{d\phi _{0}}{dx}\right) \frac{%
d\eta }{dx}+U_{0}\eta =\omega ^{2}\eta ,  \label{H1x}
\end{equation}%
where $U_{0}$ is given by
\begin{equation}
U_{0}(x)=\left. \frac{1}{f}\frac{\partial ^{2}V}{\partial \phi ^{2}}%
\right\vert _{\phi =\phi _{0}}-\left. \frac{1}{f}\frac{\partial f}{\partial
\phi }\right\vert _{\phi =\phi _{0}}\frac{d^{2}\phi _{0}}{dx^{2}}-\left.
\frac{1}{2f}\frac{\partial ^{2}f}{\partial \phi ^{2}}\right\vert _{\phi
=\phi _{0}}\left( \frac{d\phi _{0}}{dx}\right) ^{2}.
\end{equation}

\begin{figure}[!ht]
\centering
\includegraphics[height=6.5cm,width=7.5cm]{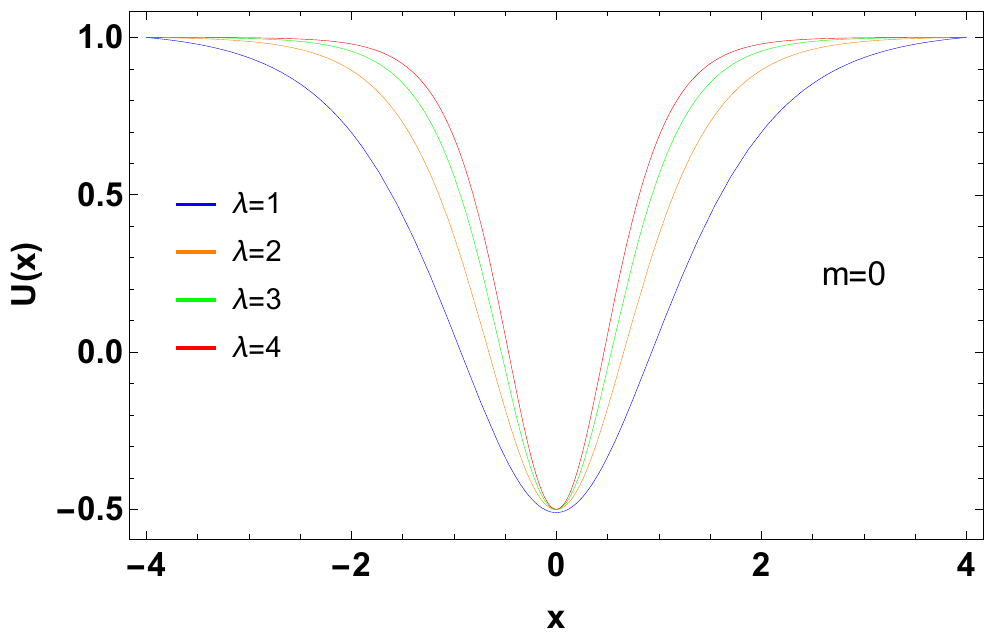} \vspace{-0.3cm}
\caption{Stability potential $U(x)$ (\ref{EEP}) keeping $m=0$, i.e., the one for the standard $\phi^4$-potential.} \label{fig6am0}
\end{figure}

The eigenvalue equation (\ref{H1x}) by means of the change of variable
\begin{equation}
\eta \left( x\right) =\frac{\psi \left( x\right) }{\sqrt{f\left( \phi
_{0}\right) }},
\end{equation}
is transformed in a Schr\"{o}dinger-like equation \cite{Vachaspati3}
\begin{equation}
\hat{H}\psi (x)=\omega ^{2}\psi (x),  \label{H111}
\end{equation}
with $\omega ^{2}$ being the eigenvalues and $\hat{H}_{0}$ is the Hamiltonian operator defined as
\begin{equation}
\hat{H}=-\frac{d^{2}}{dx^{2}}+U(x),
\end{equation}%
where the effective stability potential is
\begin{equation}
U(x)=\left. \frac{1}{f}\frac{\partial ^{2}V}{\partial \phi ^{2}}\right\vert
_{\phi =\phi _{0}}-\left. \frac{1}{2f}\frac{\partial f}{\partial \phi }%
\right\vert _{\phi =\phi _{0}}\frac{d^{2}\phi _{0}}{dx^{2}}-\left. \frac{1}{%
4f^{2}}\left( \frac{\partial f}{\partial \phi }\right) ^{2}\right\vert
_{\phi =\phi _{0}}\left( \frac{d\phi _{0}}{dx}\right) ^{2},  \label{EEP}
\end{equation}
where%
\begin{eqnarray}
\left. \frac{\partial ^{2}V}{\partial \phi ^{2}}\right\vert _{\phi =\phi
_{0}} &=&-\frac{\lambda }{\cos \left( m\pi \phi _{0}\right) }(\nu^{2}-3\phi
_{0}^{2})-\frac{2\lambda m\pi \sin \left( m\pi \phi _{0}\right) }{\cos
^{2}\left( m\pi \phi _{0}\right) }\phi _{0}(\nu^{2}-\phi _{0}^{2})  \notag \\%
[0.2cm]
&&+\frac{\lambda \left( m\pi \right) ^{2}\left[ \sin ^{2}\left( m\pi \phi
_{0}\right) +1\right] }{4\cos ^{3}\left( m\pi \phi _{0}\right) }%
(\nu^{2}-\phi _{0}^{2})^{2},
\end{eqnarray}%
and
\begin{equation}
\frac{d^{2}\phi _{0}}{dx^{2}}=\frac{\lambda }{2}\frac{m\pi \sin \left( m\pi
\phi _{0}\right) }{\cos ^{3}\left( m\pi \phi _{0}\right) }\left(
\nu^{2}-\phi _{0}^{2}\right) ^{2}-\frac{\lambda }{\cos ^{2}\left( m\pi \phi
_{0}\right) }\phi _{0}\left( \nu^{2}-\phi _{0}^{2}\right).
\end{equation}
We display the stability potential (\ref{EEP}) in Figs. \ref{fig6am0}, \ref{fig5}, and \ref{fig6}.

\begin{figure}[!ht]
\includegraphics[height=6.5cm,width=7.5cm]{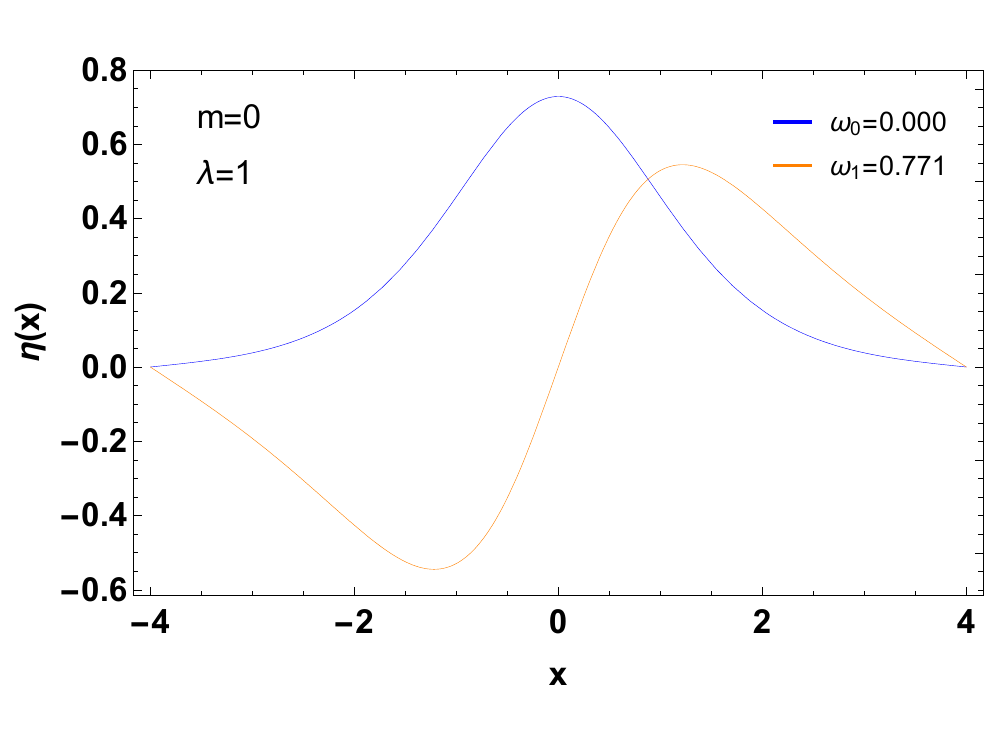}
\includegraphics[height=6.5cm,width=7.5cm]{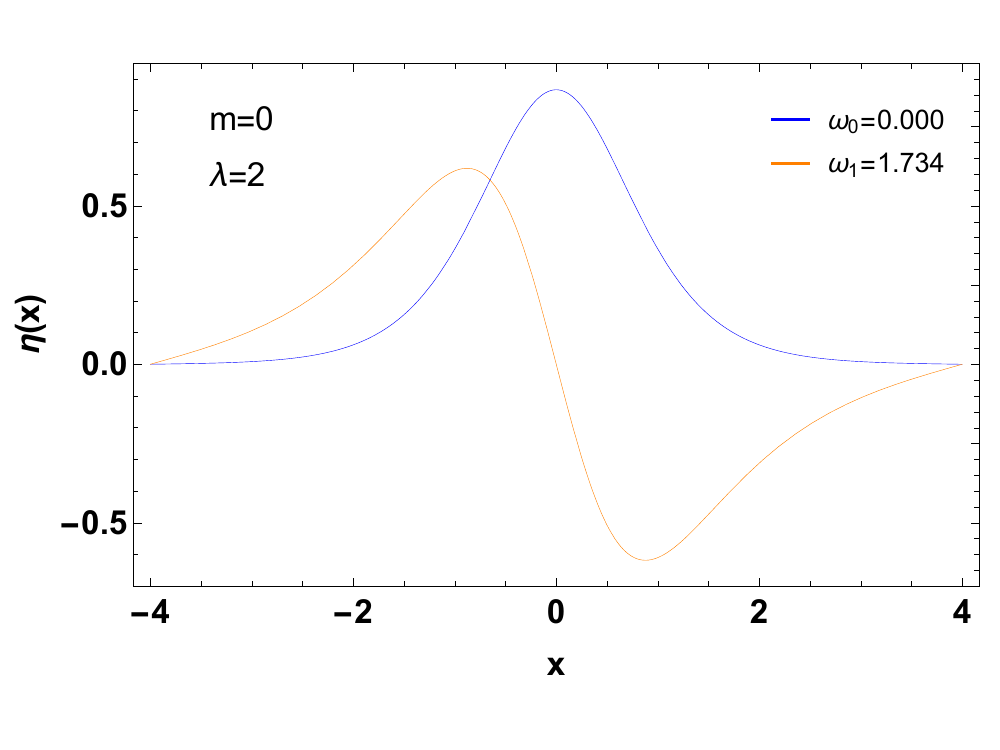}\vspace{-0.7cm}

\hspace{0.8cm} (a) \hspace{6.9cm} (b) \vspace{-0.3cm}

\includegraphics[height=6.5cm,width=7.5cm]{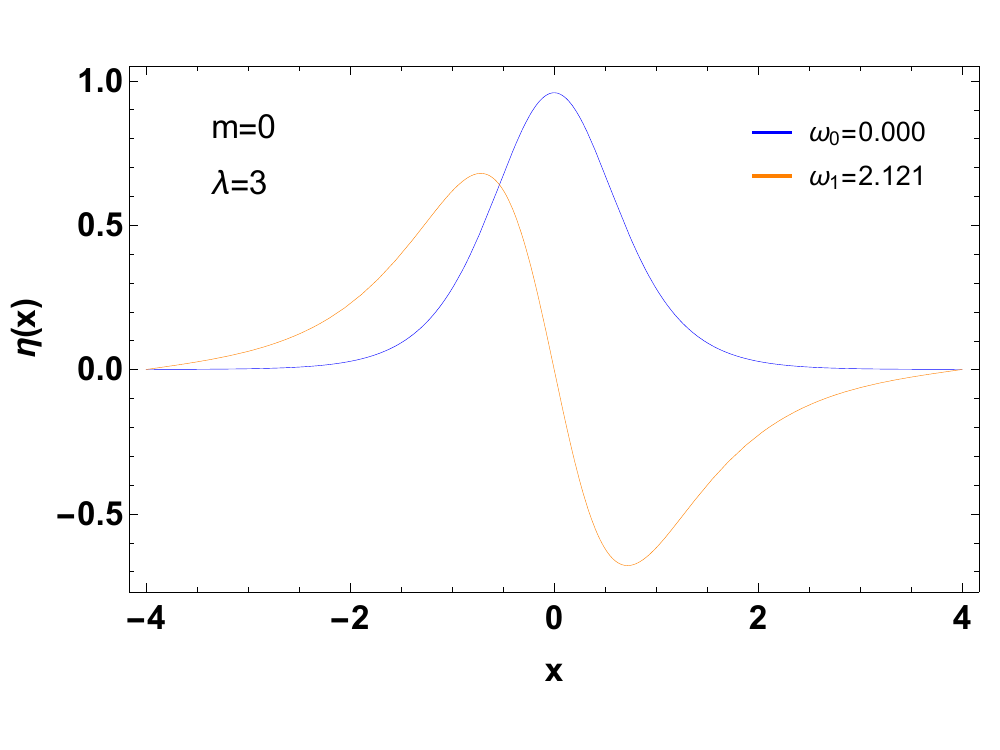} %
\includegraphics[height=6.5cm,width=7.5cm]{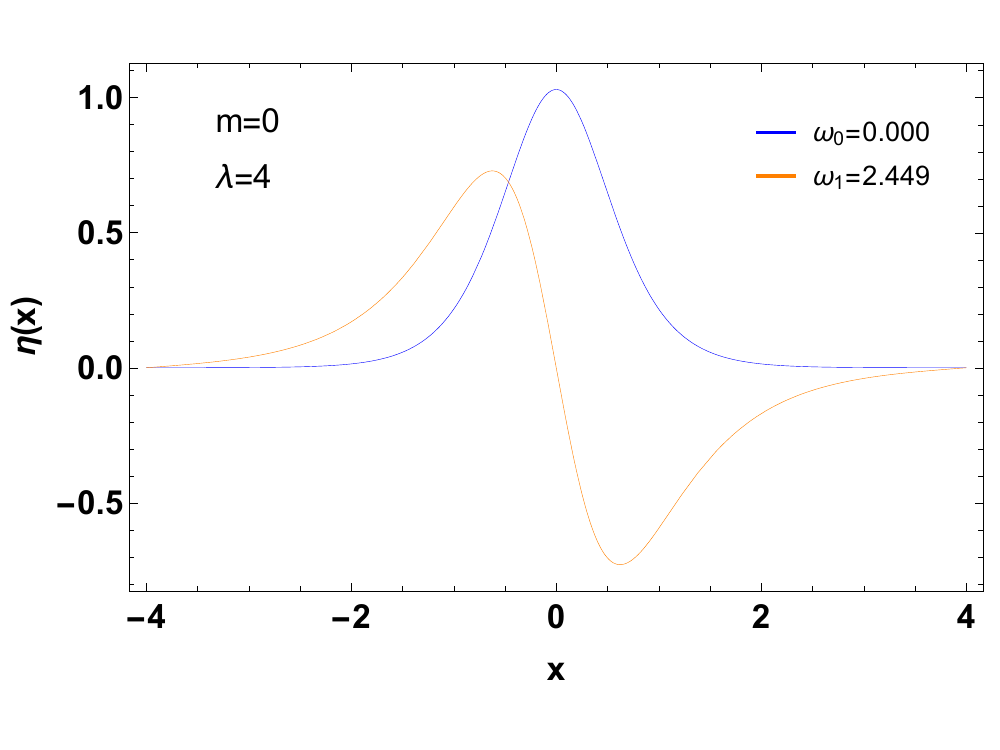}\vspace{-0.7cm}

\hspace{0.8cm} (c) \hspace{6.9cm} (d) \vspace{-0.3cm}
\caption{Translational
and vibrational modes keeping $m=0$ and varying $\lambda$, i.e., the ones for the standard $\phi^4$-potential.}
\label{fig6bm0}
\end{figure}

\begin{figure}[!ht]
\centering
\scalebox{1.05}{\includegraphics[height=6.5cm,width=7.5cm]{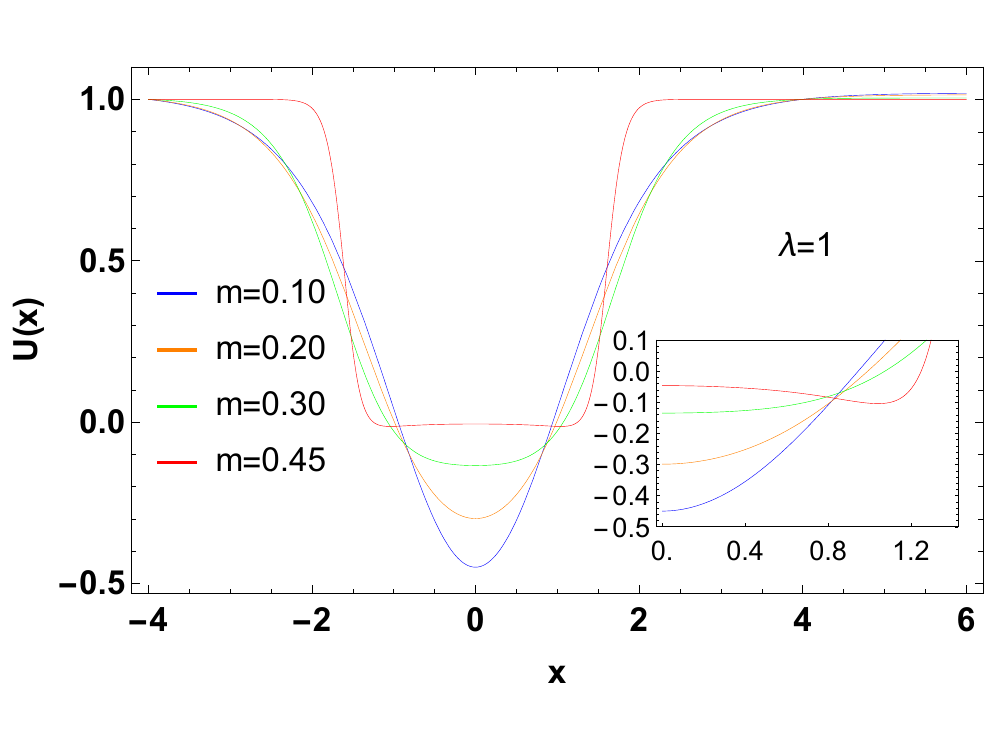}} %
\scalebox{1.05}{\includegraphics[height=6.5cm,width=7.5cm]{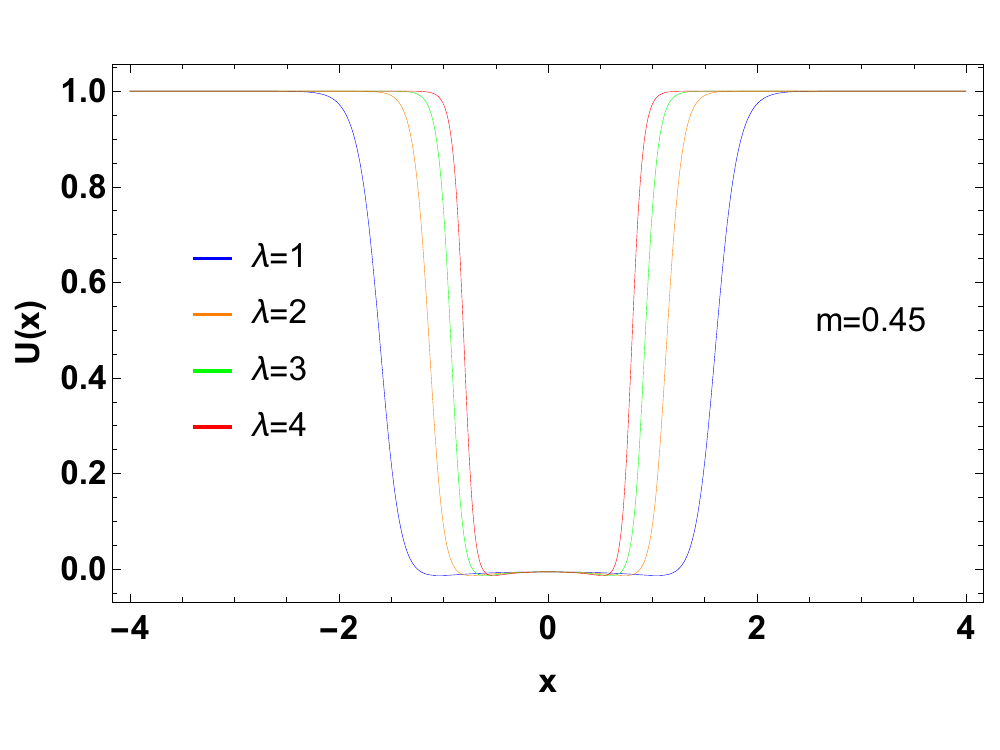}}

\hspace{0.7cm} (a) \hspace{6.8cm} (b) \vspace{-0.3cm}
\caption{(a) Stability potential $U(x)$ (\ref{EEP}) keeping $\lambda=1$  and $m$ varying. (b) Stability potential $U(x)$ (\ref{EEP}) keeping $m=0.45$ and $\lambda$
varying.}
\label{fig5}
\end{figure}

\begin{figure}[!ht]
\centering
\scalebox{1.25}{\includegraphics[height=6.5cm,width=7.5cm]{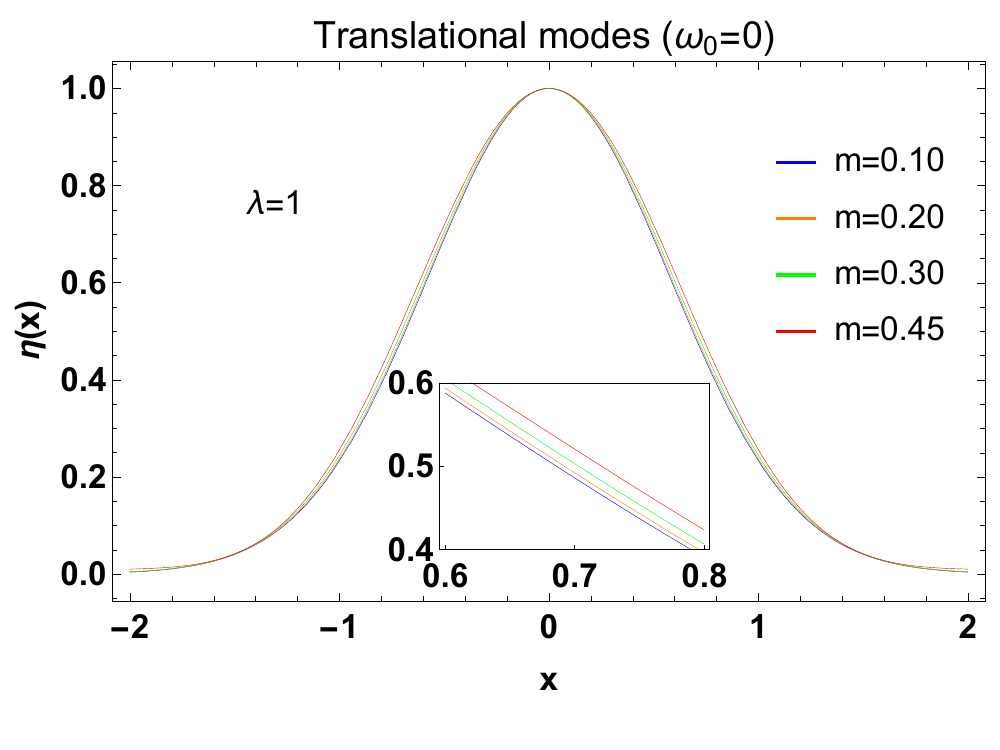}}\vspace{-0.7cm}
\caption{Translational mode, i.e., eigenfunctions $\eta(x)$ vs. $x$ for the eigenvalue null.}
\label{Zeromod}
\end{figure}

\begin{figure}[!ht]
\centering
\includegraphics[height=6.5cm,width=7.5cm]{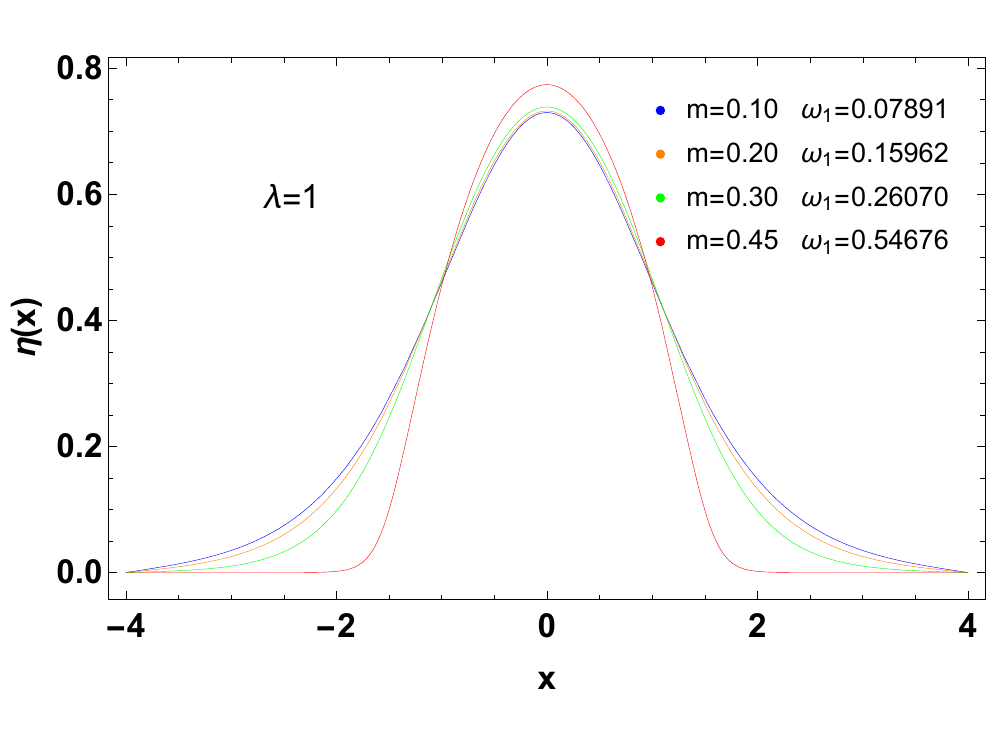}\vspace{-0.7cm}
\includegraphics[height=6.5cm,width=7.5cm]{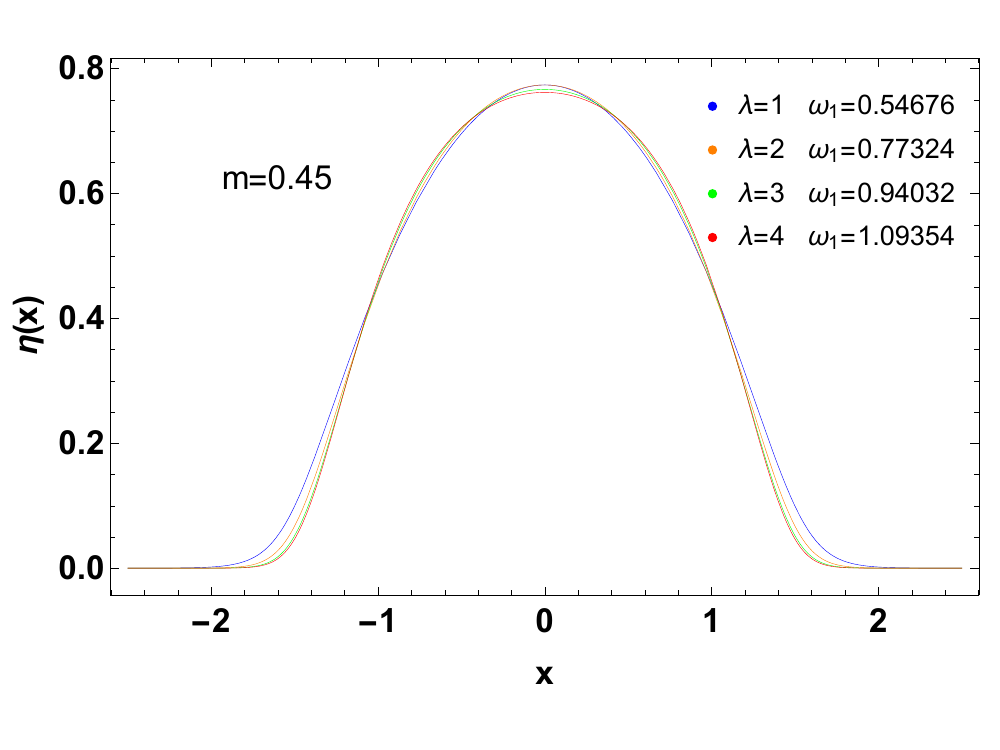}\vspace{-0.7cm}

\hspace{0.65cm} (a) \hspace{6.9cm} (b) \vspace{-0.3cm}
\caption{(a)
Eigenfunctions of Eq. (\ref{H1x}) representing the ground states (first vibrational modes) for some values of $m$ and $\lambda=1$. (b) Eigenfunctions of Eq. (\ref{H1x}) representing the ground
states (first vibrational modes) for some values of $\lambda$ and $m=0.45$}.
\label{fig6}
\end{figure}

Utilizing the finite element method, we investigate the model eigenstates \cite{Belendryasova, Hildebrand}. We first illustrate in Figs. \ref{fig6am0} and \ref{fig6bm0} the stability potential, and corresponding eigenfunctions of the case $m=0$ (i.e., those of the standard $\phi^4$-theory). In this scenario, the model admits a translational mode (or zero mode, i.e., the eigenfunction with null eigenvalue, $\omega=0$) and one vibrational eigenstate ($\omega^2>0$). Besides, one notes through numerical inspection that this spectrum remains unchanged for sufficiently small $m$ values, but starting from sufficiently larger $m$ values, one translational mode and several vibrational modes emerge. The stability potential, translational mode, and first vibrational modes, along with the associated eigenvalues, are depicted in Figs. \ref{fig5}[(a) and (b)], \ref{Zeromod}, and \ref{fig6}[(a) and (b)], respectively. Notably, the presence of the translational and vibrational modes suggests the occurrence of resonances in the scattering of configurations with opposing topological charges.

\begin{figure}[!ht]
\centering
\includegraphics[height=6.5cm,width=7cm]{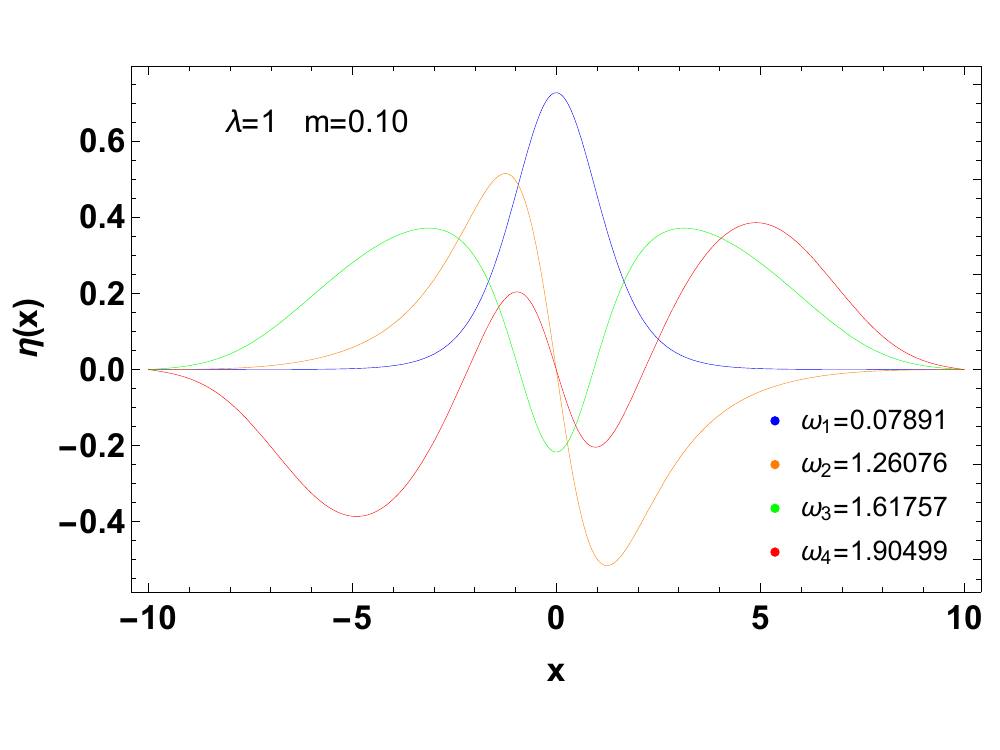} %
\includegraphics[height=6.5cm,width=7cm]{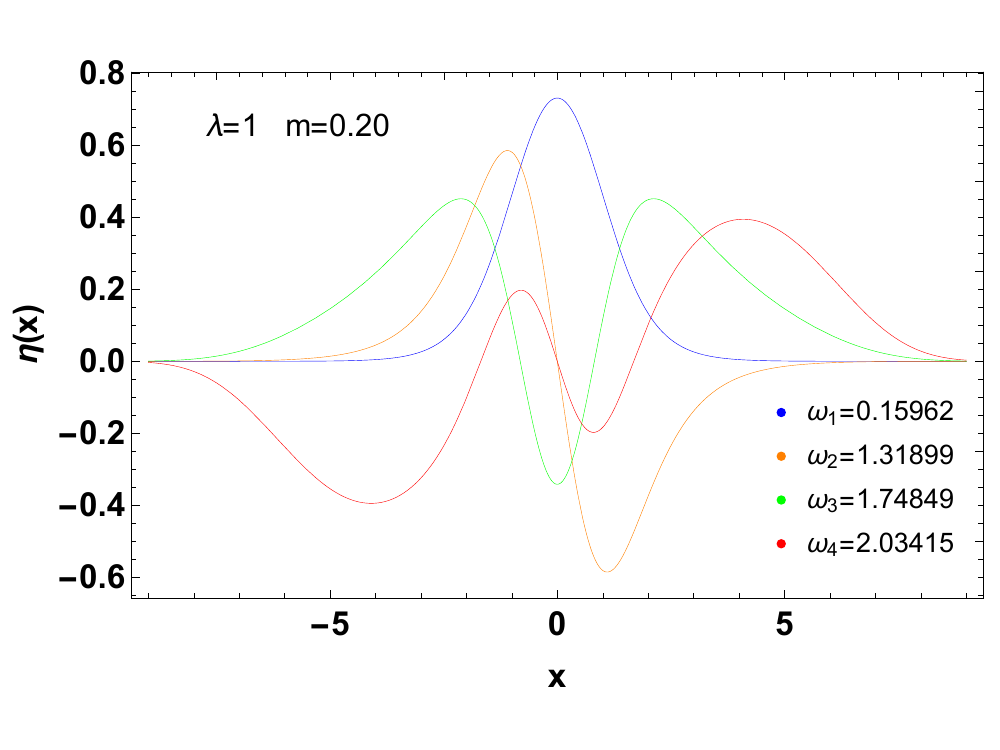}\vspace{-0.7cm}

\hspace{0.8cm} (a) \hspace{6.4cm} (b) \vspace{-0.2cm}

\includegraphics[height=6.5cm,width=7cm]{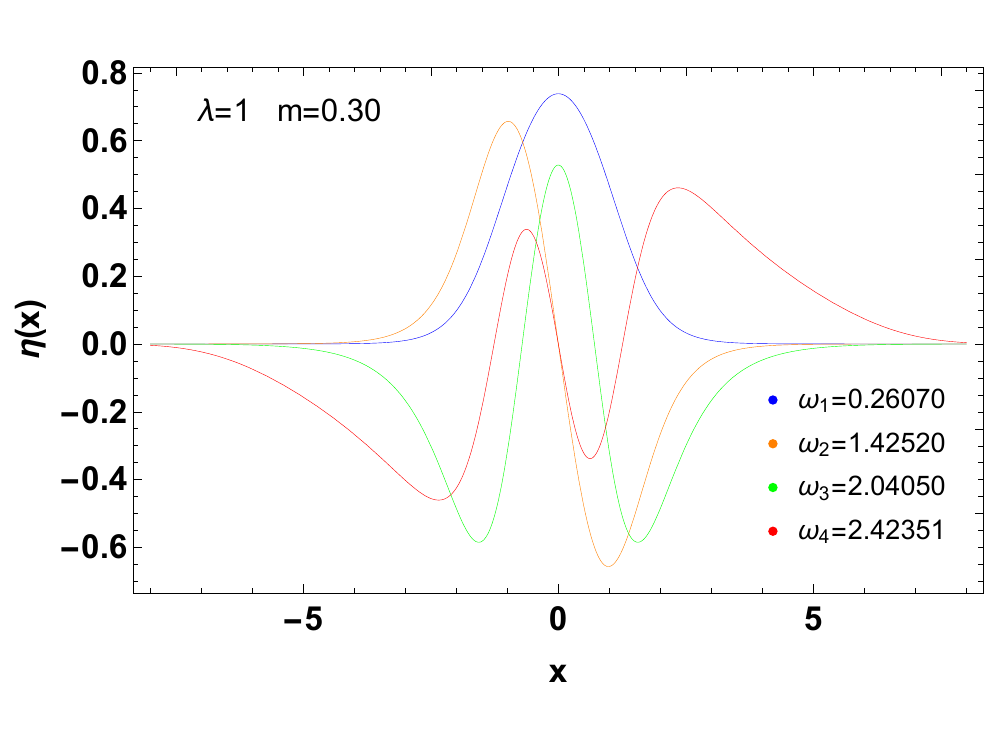} %
\includegraphics[height=6.5cm,width=7cm]{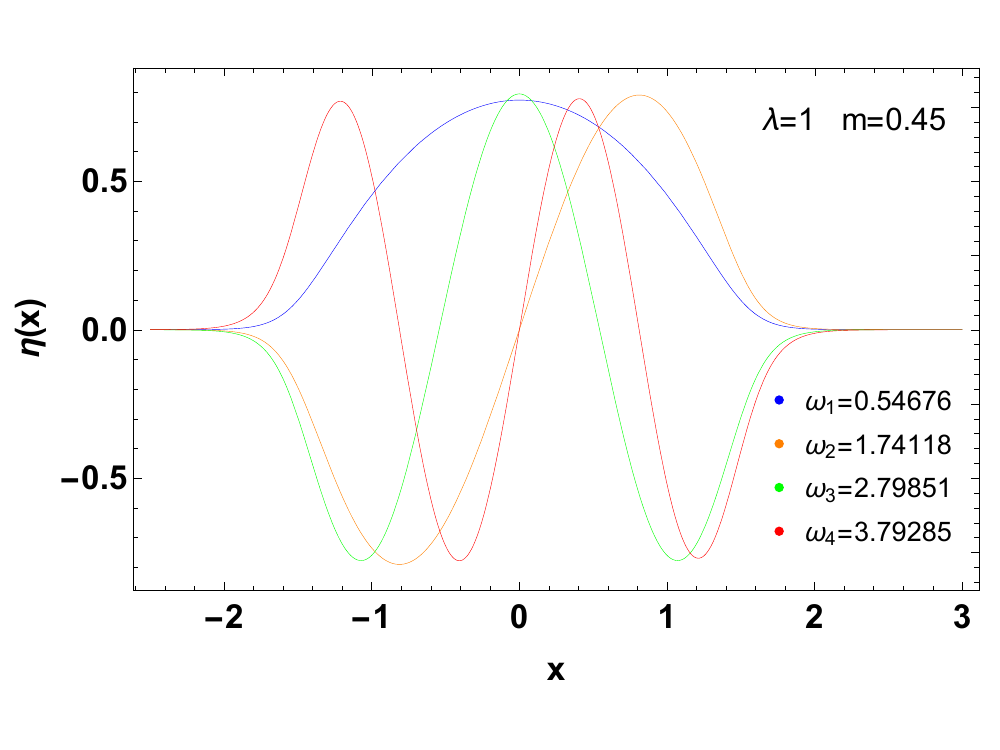} \vspace{-0.7cm}

\hspace{0.7cm} (c) \hspace{6.4cm} (d) \vspace{-0.3cm}
\caption{First vibrational modes solutions of the Eq. (\ref{H1x}), i.e., $n=1$, $2$, $3$, and $4$. One assumes for all cases $\lambda=1$. (a) Eigenfunctions when $m=0.10$. (b) Eigenfunctions when $m=0.20$. (c) Eigenfunctions when $m=0.30$. (d) Eigenfunctions when $m=0.45$.} \label{fig7}
\end{figure}

The figures \ref{fig7}[(a), (b), (c), and (d)] show the first vibrational modes solutions of the Eq. (\ref{H1x}) calculated with $m$ varying and $\lambda=1$ remaining constant, albeit with a more localized profile, when $m$ grows. Furthermore, in Tables I(a) and I(b), we summarize our results for the eigenvalues of the first eigenstates, where the ones in Table I(a) correspond to the Fig. \ref{fig7}.

\begin{table}[!ht]
\caption{(a) First vibrational states when $\lambda=1$. (b) First vibrational states when $m=0.45$.} \vspace{0.4cm}

\centering \hspace{-8.5cm}
\resizebox{7.5cm}{8cm}{
    \begin{tabular}{|c|c|c|}\hline\hline
    $m$-parameter    &  $n$-th eigenstate & $\omega_n$-eigenvalue  \\ \hline \hline
    \multirow{5}{*}{0.10}   &  1 &  0.07891 \\ \cline{2-3}
    &  2 &  1.26076 \\ \cline{2-3}
    &  3 &  1.61757 \\ \cline{2-3}
    &  4 &  1.90489 \\ \cline{2-3}
    &  5 & 2.57862  \\ \hline
    \multirow{5}{*}{0.20}   &  1 &  0.15962 \\ \cline{2-3}
    &  2 &  1.31899 \\ \cline{2-3}
    &  3 & 1.74849  \\ \cline{2-3}
    &  4 &  2.03415 \\ \cline{2-3}
    &  5 &  2.66395 \\ \hline
    \multirow{5}{*}{0.30} &  1 & 0.26070 \\ \cline{2-3}
    &  2 & 1.42520 \\ \cline{2-3}
    &  3 & 2.04050 \\ \cline{2-3}
    &  4 &  2.42351 \\ \cline{2-3}
    &  5 &  2.95127 \\ \hline
    \multirow{5}{*}{0.45} &  1 & 0.54676 \\ \cline{2-3}
    &  2 &  1.74118 \\ \cline{2-3}
    &  3 & 2.79851  \\ \cline{2-3}
    &  4 & 3.79285 \\ \cline{2-3}
    &  5 & 4.74162  \\ \hline
    \end{tabular}}

\vspace{-15.85cm} \hspace{8cm}
\resizebox{7.5cm}{8cm}{
    \begin{tabular}{|c|c|c|}\hline\hline
    $\lambda$-parameter    &  $n$-th eigenstate & $\omega_n$-eigenvalue  \\ \hline \hline
    \multirow{5}{*}{1}   &  1 &  0.54676 \\ \cline{2-3}
    &  2 &  1.74118 \\ \cline{2-3}
    &  3 & 2.79851  \\ \cline{2-3}
    &  4 & 3.79285  \\ \cline{2-3}
    &  5 &  4.74162 \\ \hline
    \multirow{5}{*}{2}   &  1 &  0.77324 \\ \cline{2-3}
    &  2 &  2.46240 \\ \cline{2-3}
    &  3 &  3.95770 \\ \cline{2-3}
    &  4 &  5.36391 \\ \cline{2-3}
    &  5 & 6.70566  \\ \hline
    \multirow{5}{*}{3} &  1 & 0.94032 \\ \cline{2-3}
    &  2 & 3.01581 \\ \cline{2-3}
    &  3 &  4.84717 \\ \cline{2-3}
    &  4 &  6.56942 \\ \cline{2-3}
    &  5 &  8.21272 \\ \hline
    \multirow{5}{*}{4} &  1 & 1.09354 \\ \cline{2-3}
    &  2 & 3.48236 \\ \cline{2-3}
    &  3 &  5.59703 \\ \cline{2-3}
    &  4 &  7.58571 \\ \cline{2-3}
    &  5 &  9.48324  \\ \hline
    \end{tabular}}\vspace{0.5cm}

\hspace{-0.2cm} (a) \hspace{7.45cm} (b) \vspace{-0.1cm}
\end{table}

The results show the existence of translational and vibrational modes, indicating that when configurations tend towards a compact-like profile, the resonance phenomenon will occur in kink-antikink scattering. Similarly, one emphasizes that experimental studies on kink collisions have also observed resonance phenomena even in systems without vibrational modes \cite{Belendryasova}. Furthermore, we are aware of two possible collision types for the system: the first produces energy emission with the presence of the resonance phenomenon, and the second is a collision in which the kink interactions form bions along the time evolution and energy dissipation \cite{Dusuel, Dorey1}.

In light of these hypotheses, the question arises: how will the collision of our kink and antikink structures behave? To address this inquiry, we present the study of these collisions below.

\begin{figure}[!ht]
\centering
\includegraphics[width=8.5cm]{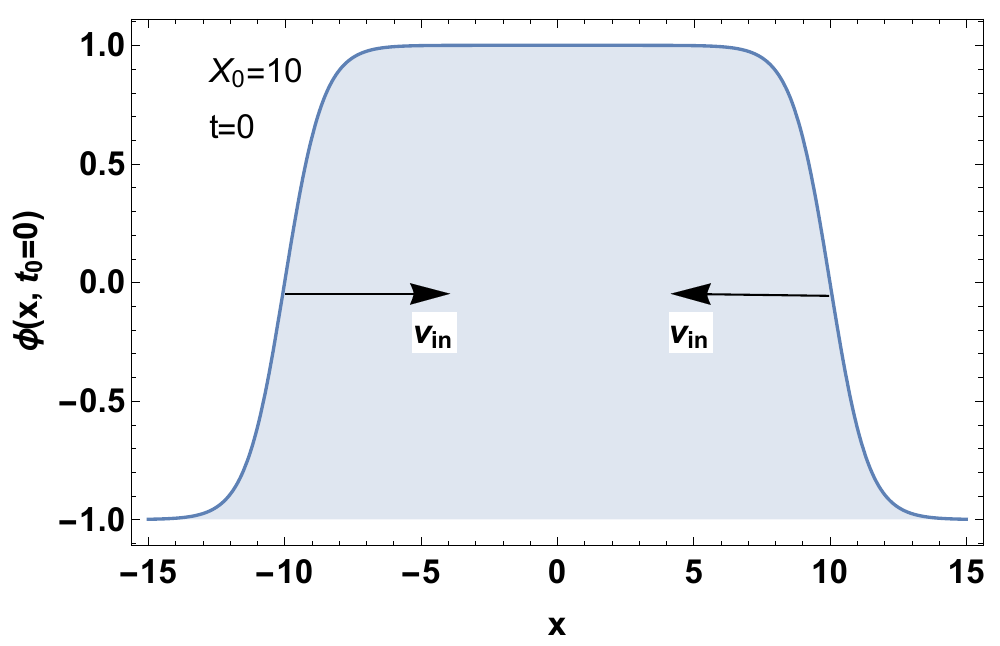} \vspace{-0.3cm}
\caption{Illustration of kink-antikink structures on the imminence of the collision when $X_0=10$, $\nu=1$, and $t=0$.} \label{fig8}
\end{figure}

\subsection{Scattering of structures}

Let us now investigate the scattering of structures identified in the previous section. In pursuit of our objective, we perform a numerical analysis of collisions involving kink-antikink configurations when these structures, as $m$ increases, transform into more compacted solutions. To study the scattering process, the equation of motion (\ref{EOM111}) is employed, with the
condition \cite{Vachaspati,Belendryasova,Campos}, \begin{equation}
\phi (x,t)=\phi _{K}[\gamma (x+X_{0}-v_{\text{in}}t)]+\phi _{\bar{K}}[\gamma (x-X_{0}+v_{\text{in}}t)]-c_{0},  \label{scattering}
\end{equation}%
(see Fig. \ref{fig8}) where the indices $K$ and $\bar{K}$ describe, respectively, kink and antikink solutions. The parameter $X_{0}$ denotes the initial position of the structures, which, for convenience, one assumes $X_{0}=10$. Furthermore, $v_{\text{in}}$ represents the
initial velocity of the configurations, and $\gamma$ is the Lorentz factor. Finally, we adopt the parameter $c_{0}=1$ to adjust the collision boundary. Thus, we will study the collisions between kink-antikink configurations that interpolate between the states $(-1,1,-1)$. For numerical analysis, we employ the finite element method with second-order discretization for temporal and spatial coordinates with steps of $10^{-3}$.

\begin{figure}[!ht]
\nonumber
\centering
\includegraphics[height=4cm,width=5cm]{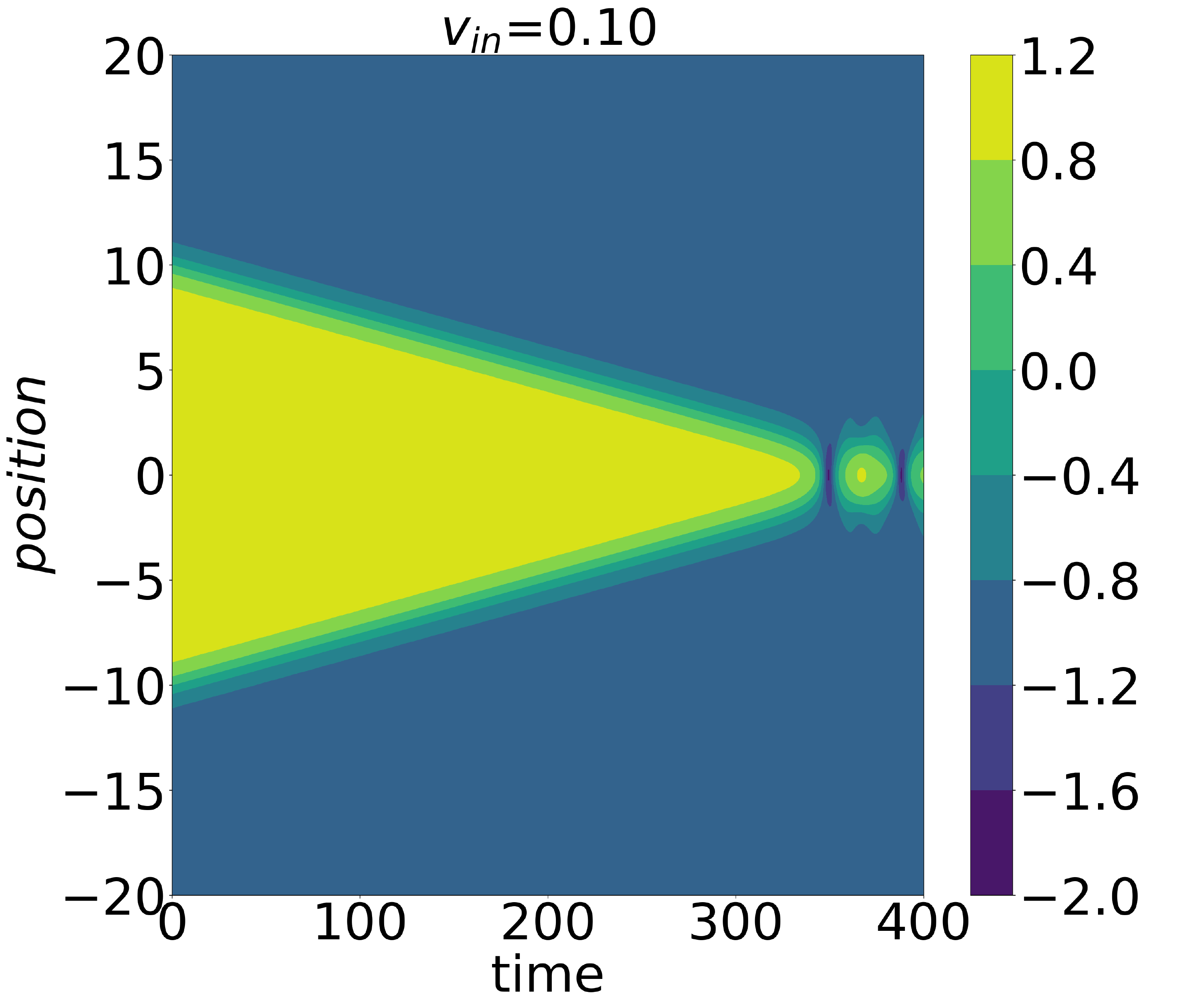} %
\includegraphics[height=4cm,width=5cm]{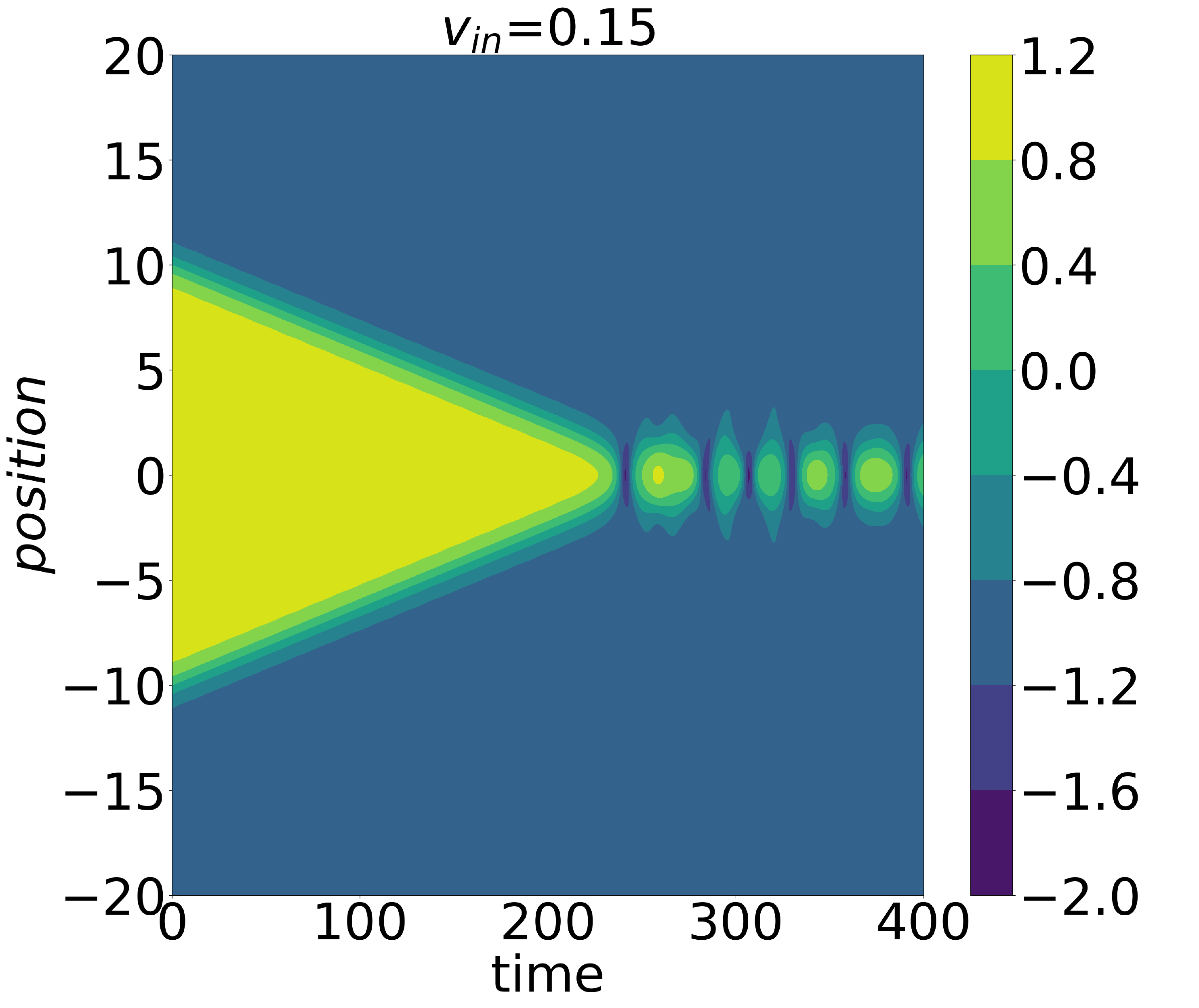} %
\includegraphics[height=4cm,width=5cm]{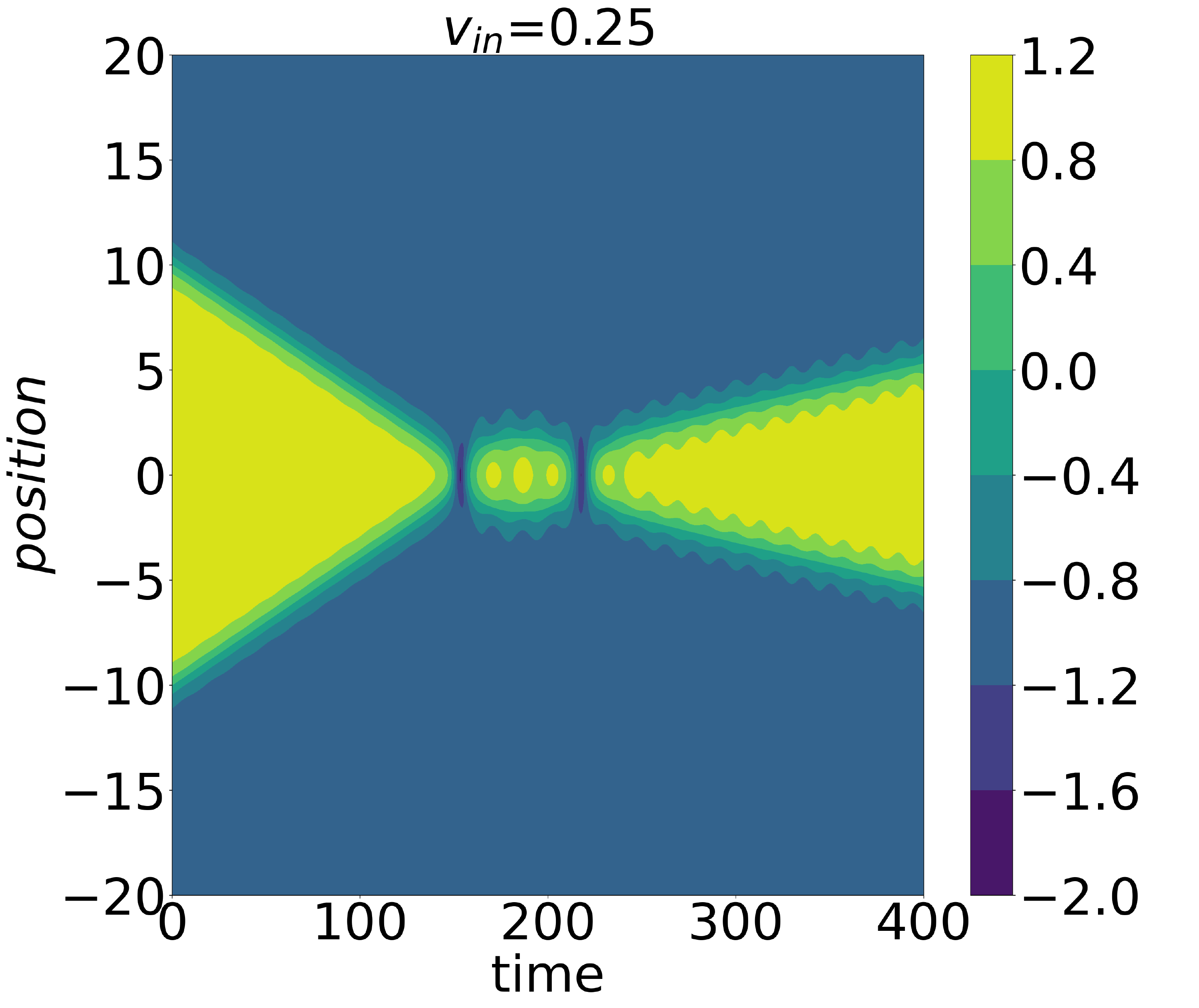}

\vspace{-0.2cm} (a) \vspace{0.3cm}

\includegraphics[height=4cm,width=5cm]{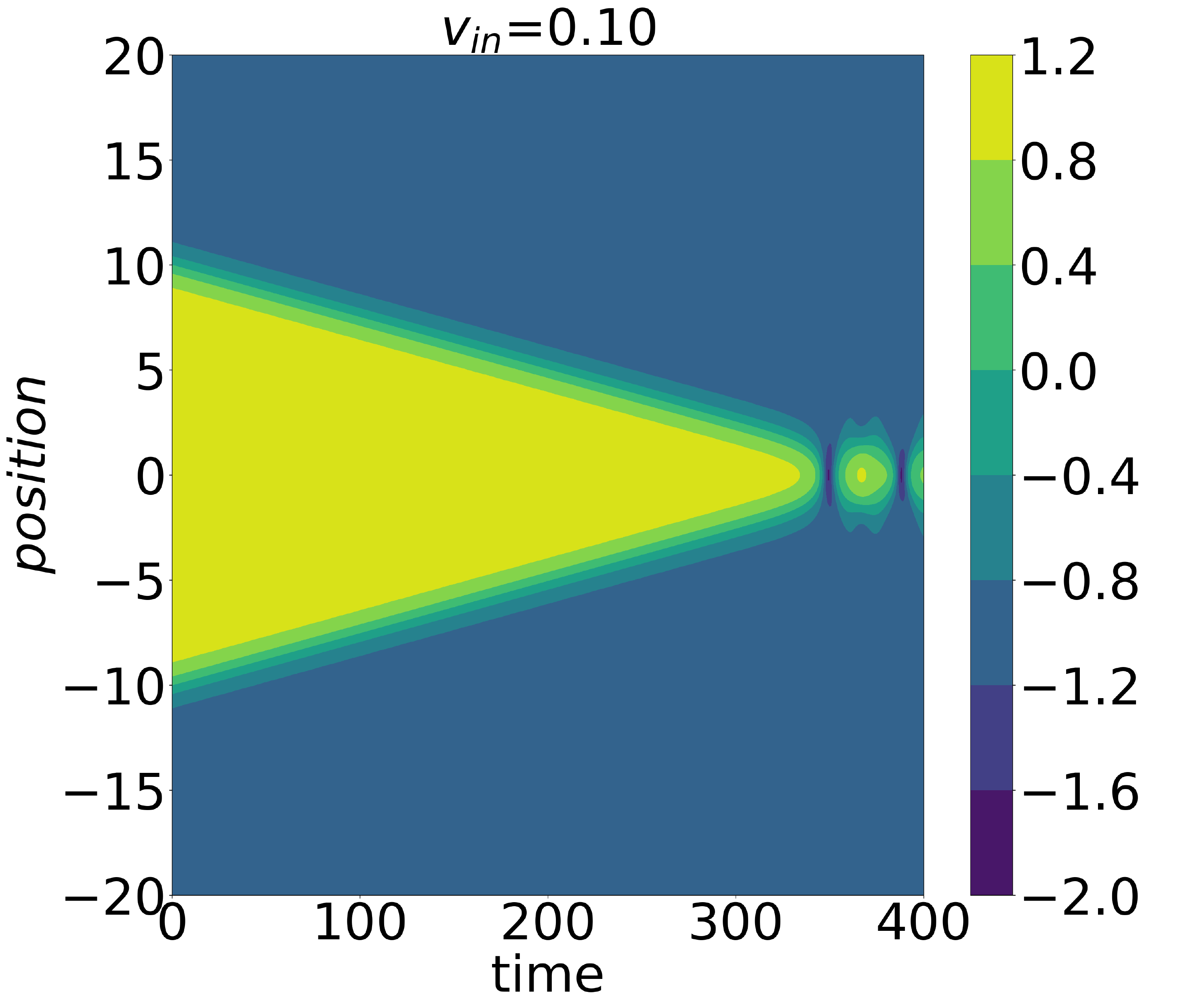} %
\includegraphics[height=4cm,width=5cm]{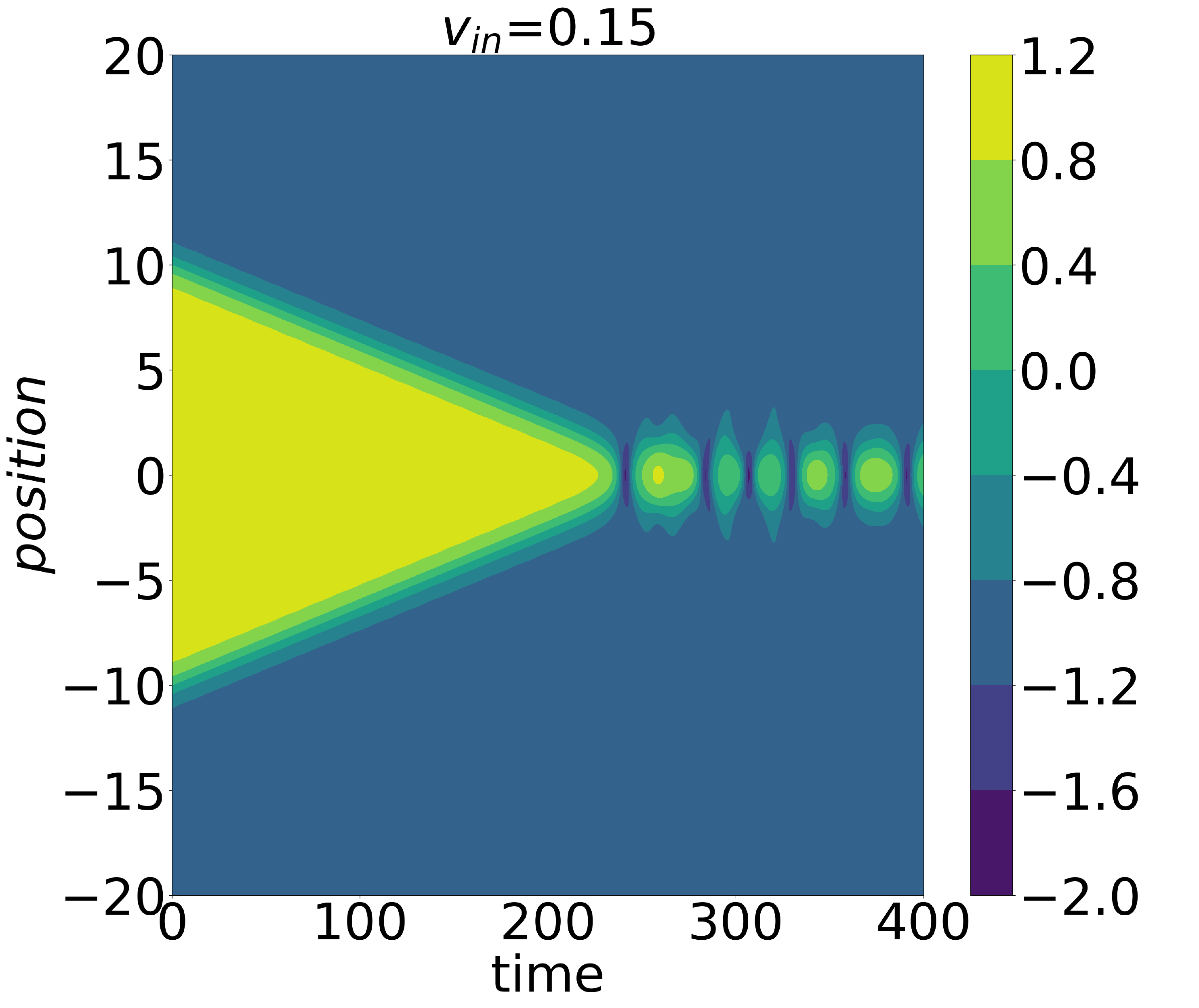} %
\includegraphics[height=4cm,width=5cm]{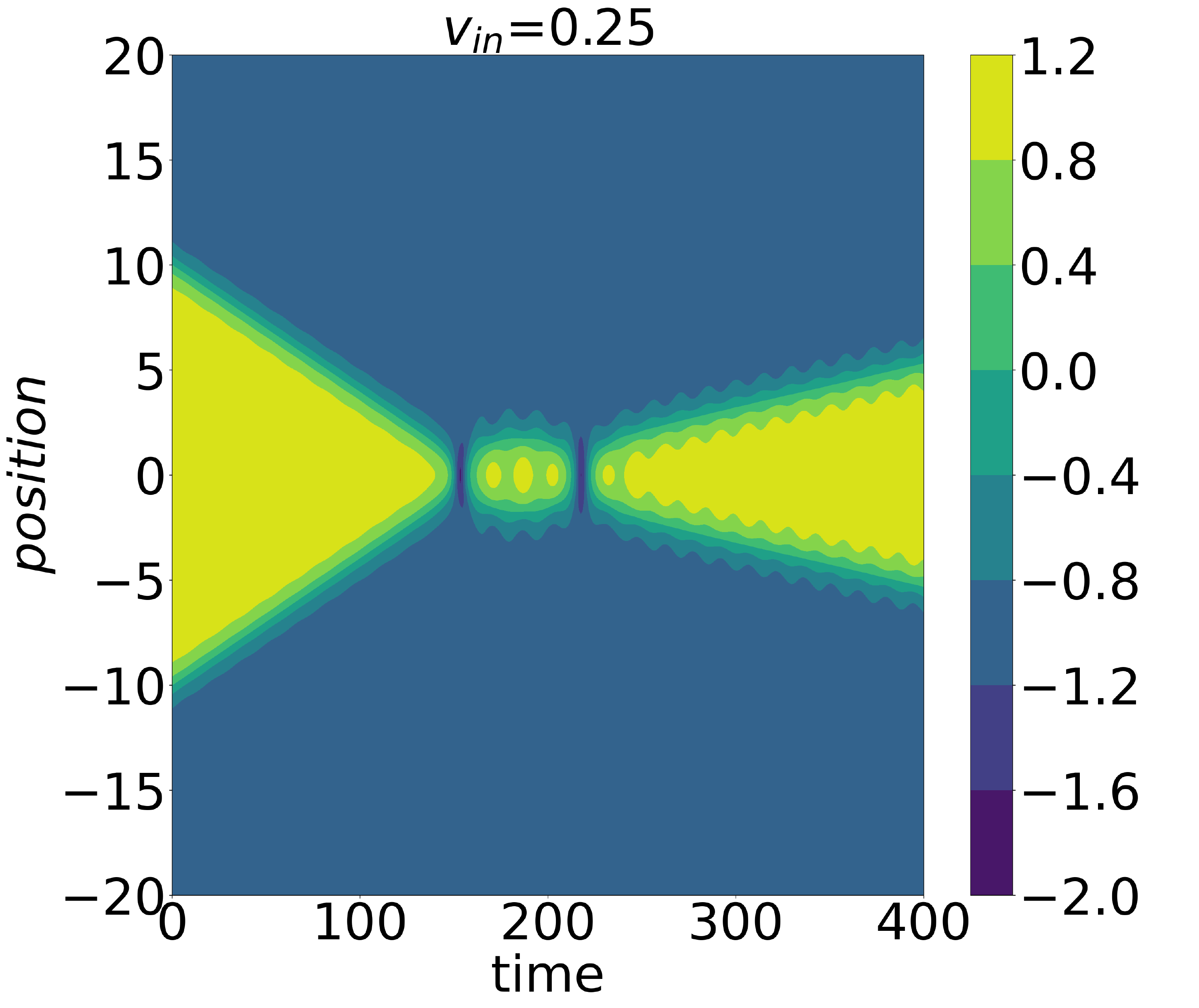}

\vspace{-0.2cm} (b) \vspace{0.3cm}

\includegraphics[height=4cm,width=5cm]{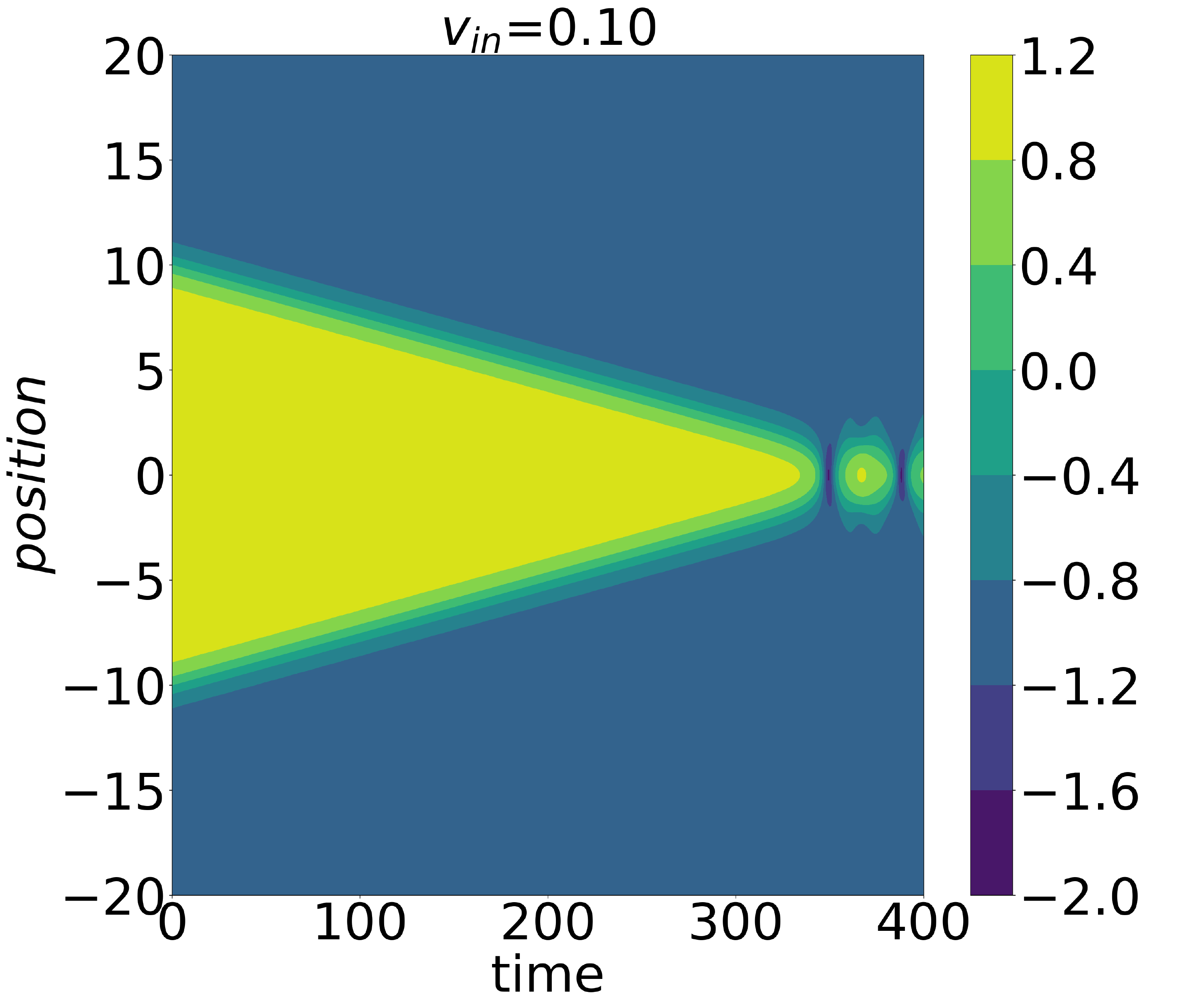} %
\includegraphics[height=4cm,width=5cm]{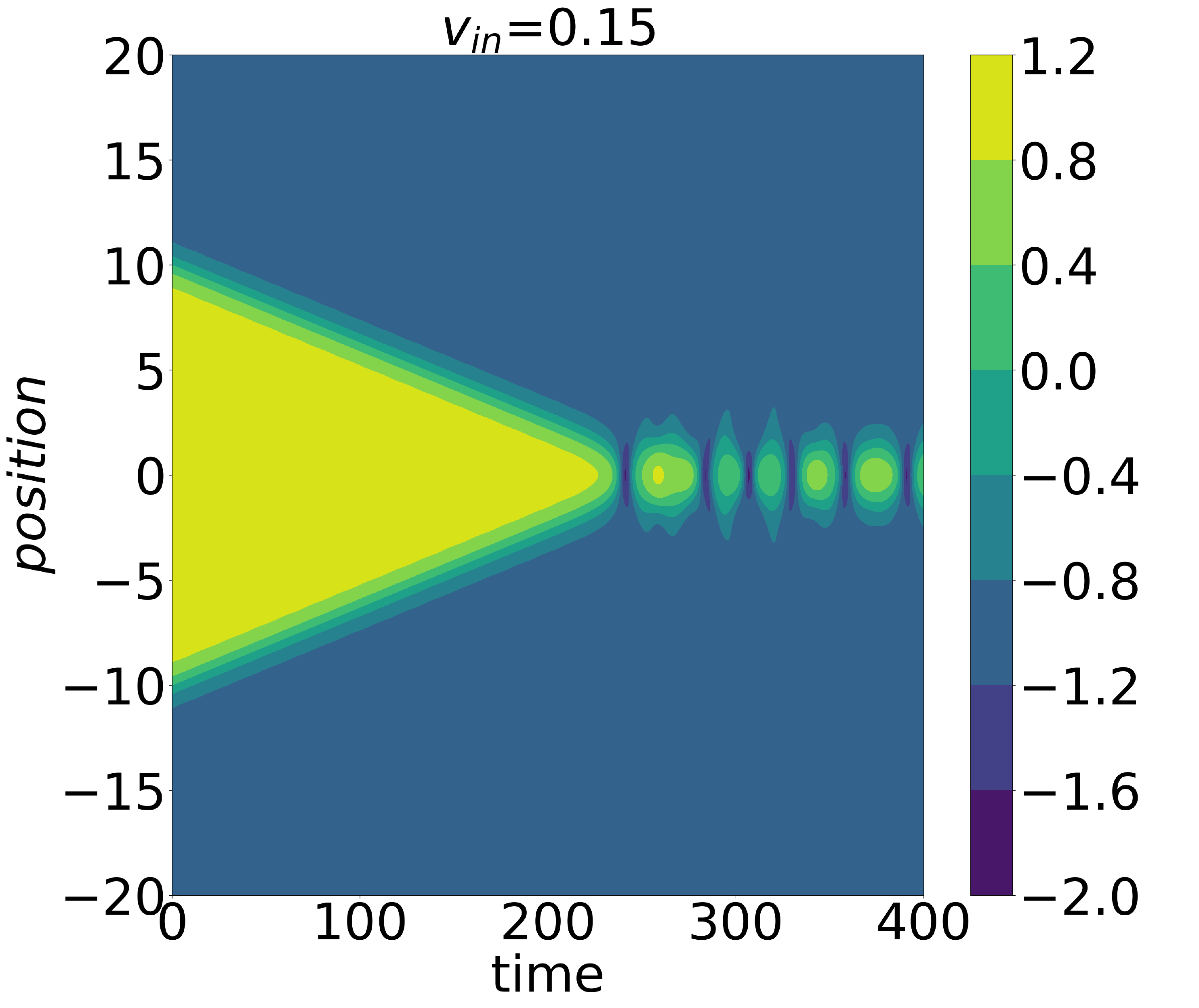} %
\includegraphics[height=4cm,width=5cm]{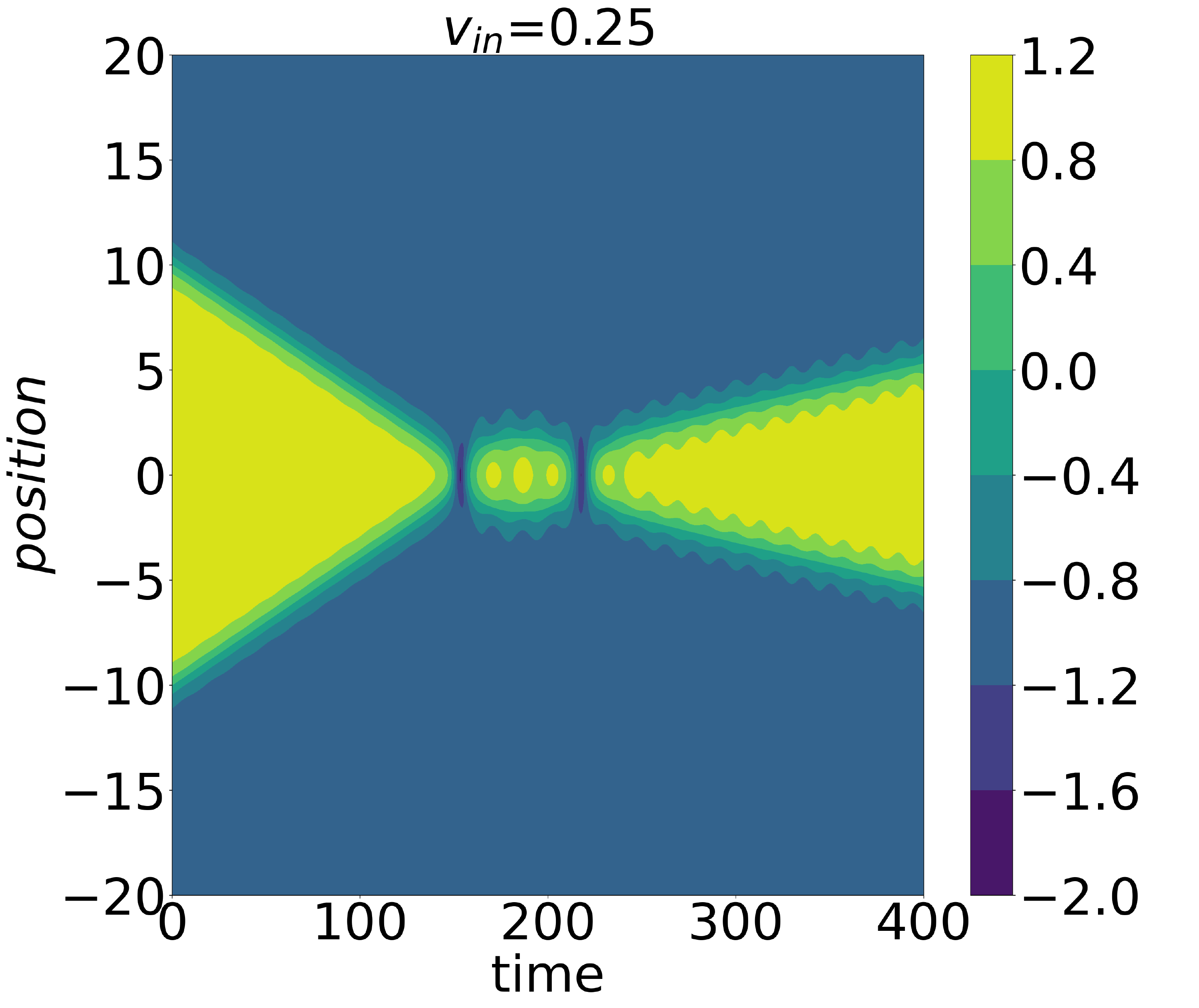}

\vspace{-0.2cm} (c) \vspace{0.3cm}

\includegraphics[height=4cm,width=5cm]{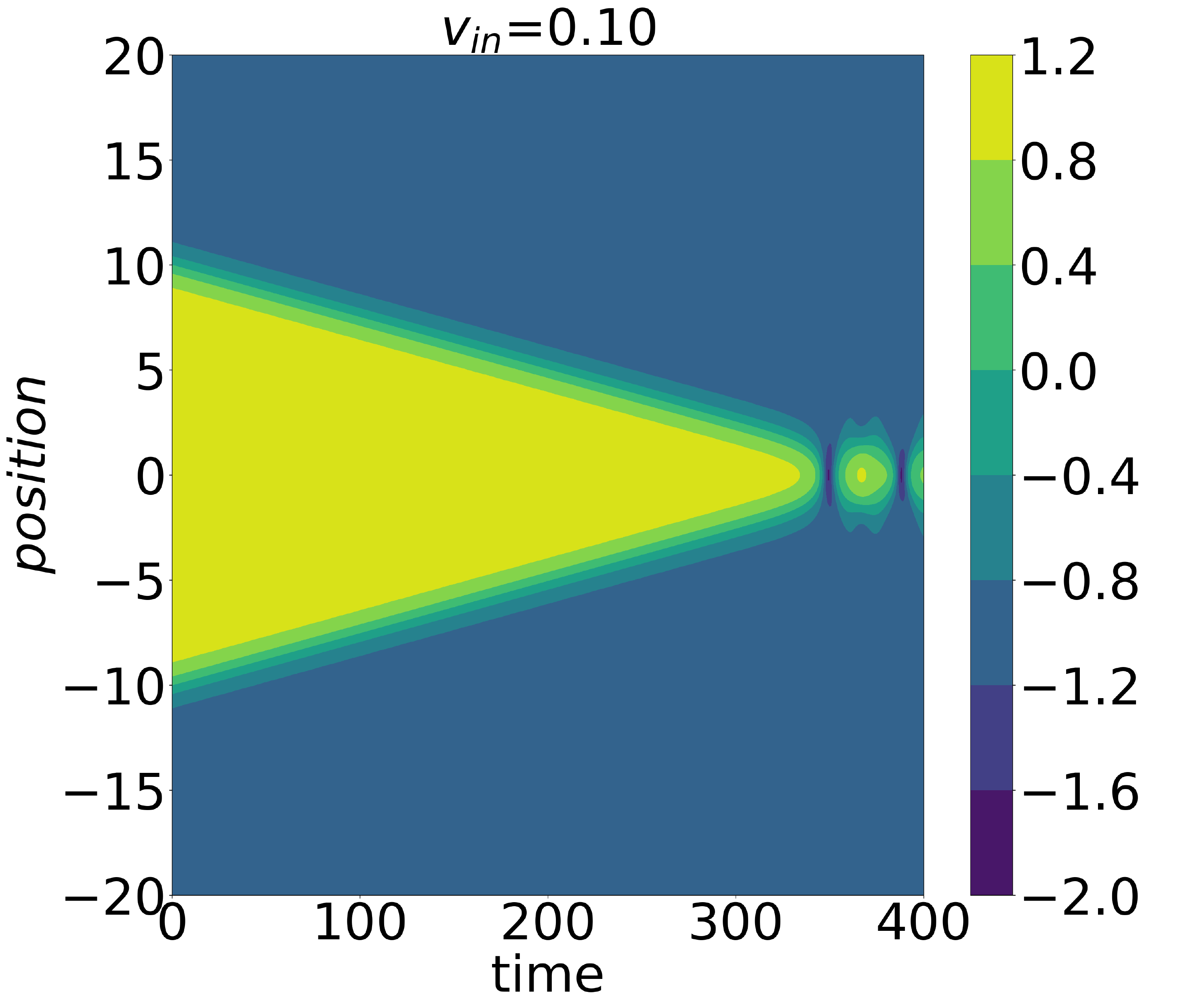} %
\includegraphics[height=4cm,width=5cm]{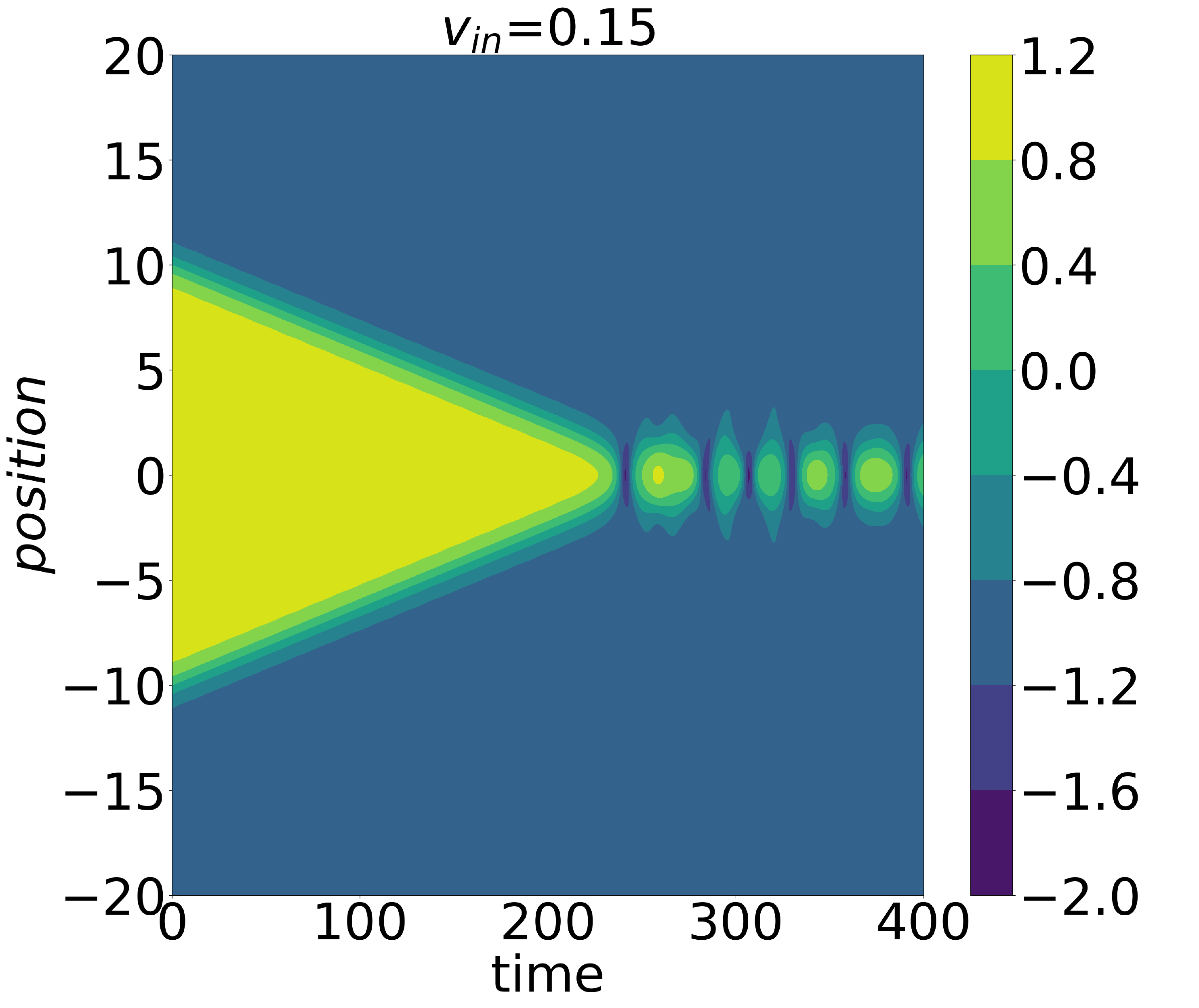} %
\includegraphics[height=4cm,width=5cm]{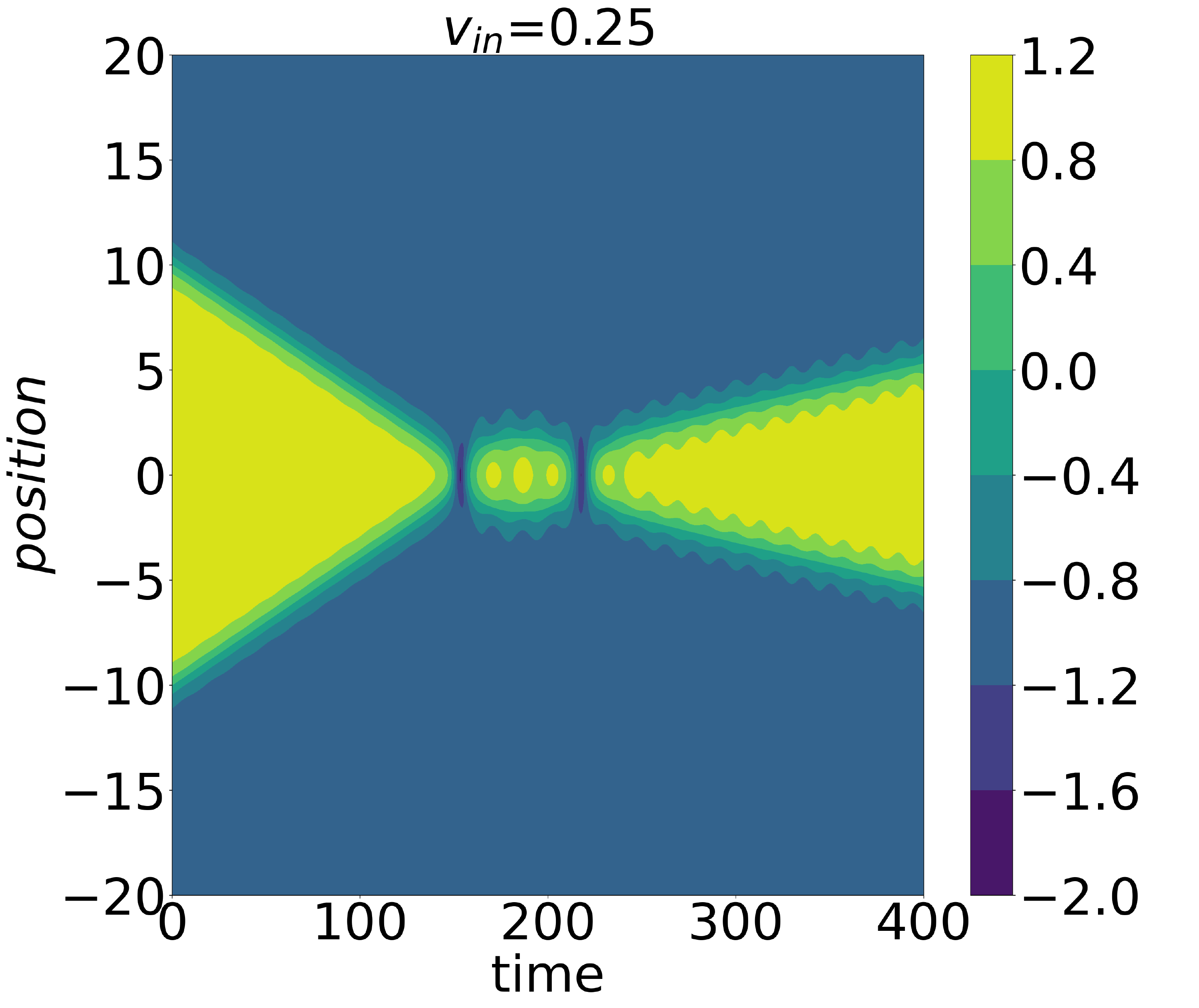}

\vspace{-0.2cm} (d) \vspace{-0.3cm}
\caption{(a) Collision $\lambda=1$ and
$m=0.10$. (b)Collision $\lambda=1$ and $m=0.20$. (c) Collision $\lambda=1$
and $m=0.30$. (d) Collision $\lambda=1$ and $m=0.45$.} \label{fig9}
\end{figure}

\begin{figure}[!ht]
\nonumber
\centering
\includegraphics[height=4cm,width=5cm]{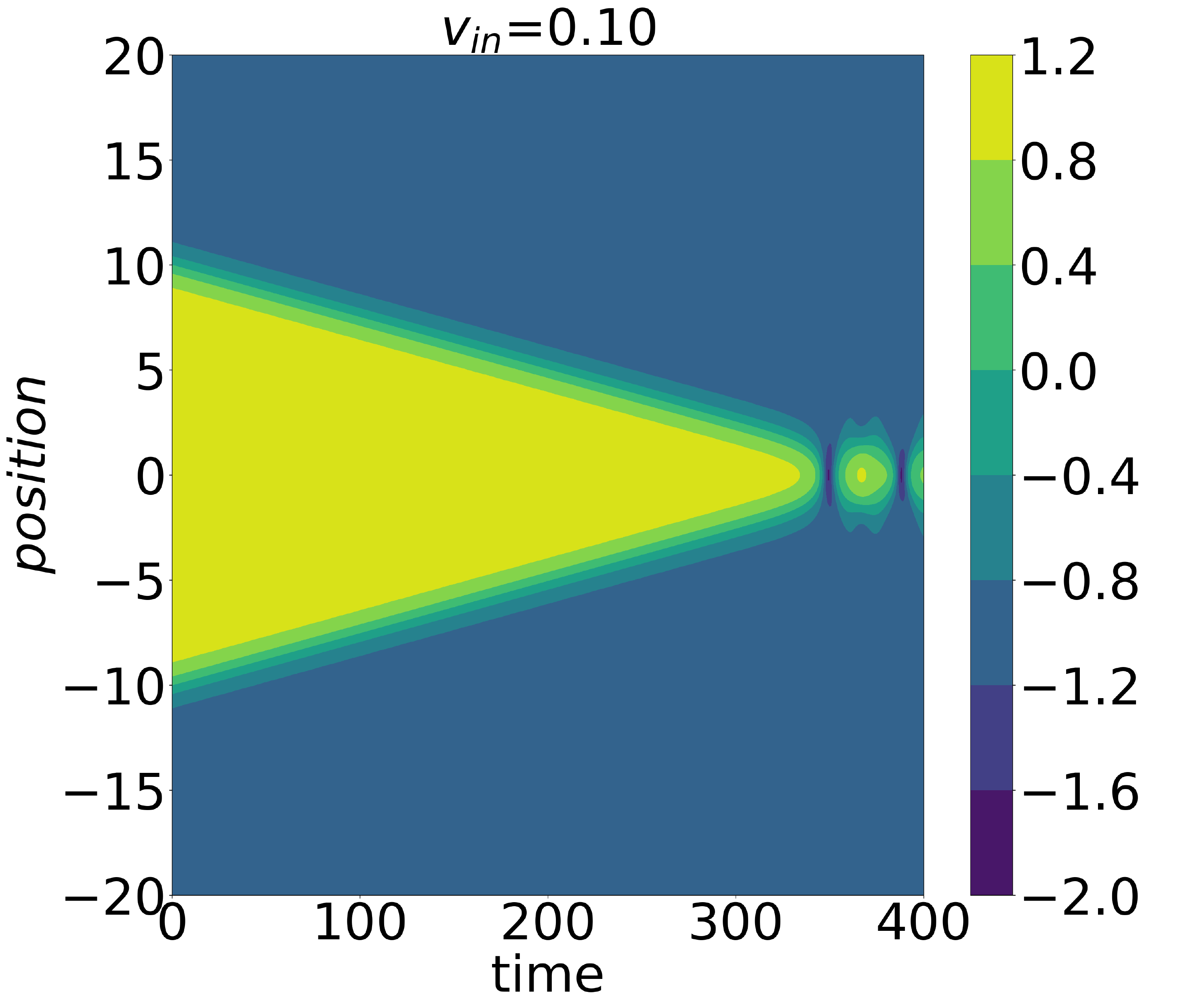} %
\includegraphics[height=4cm,width=5cm]{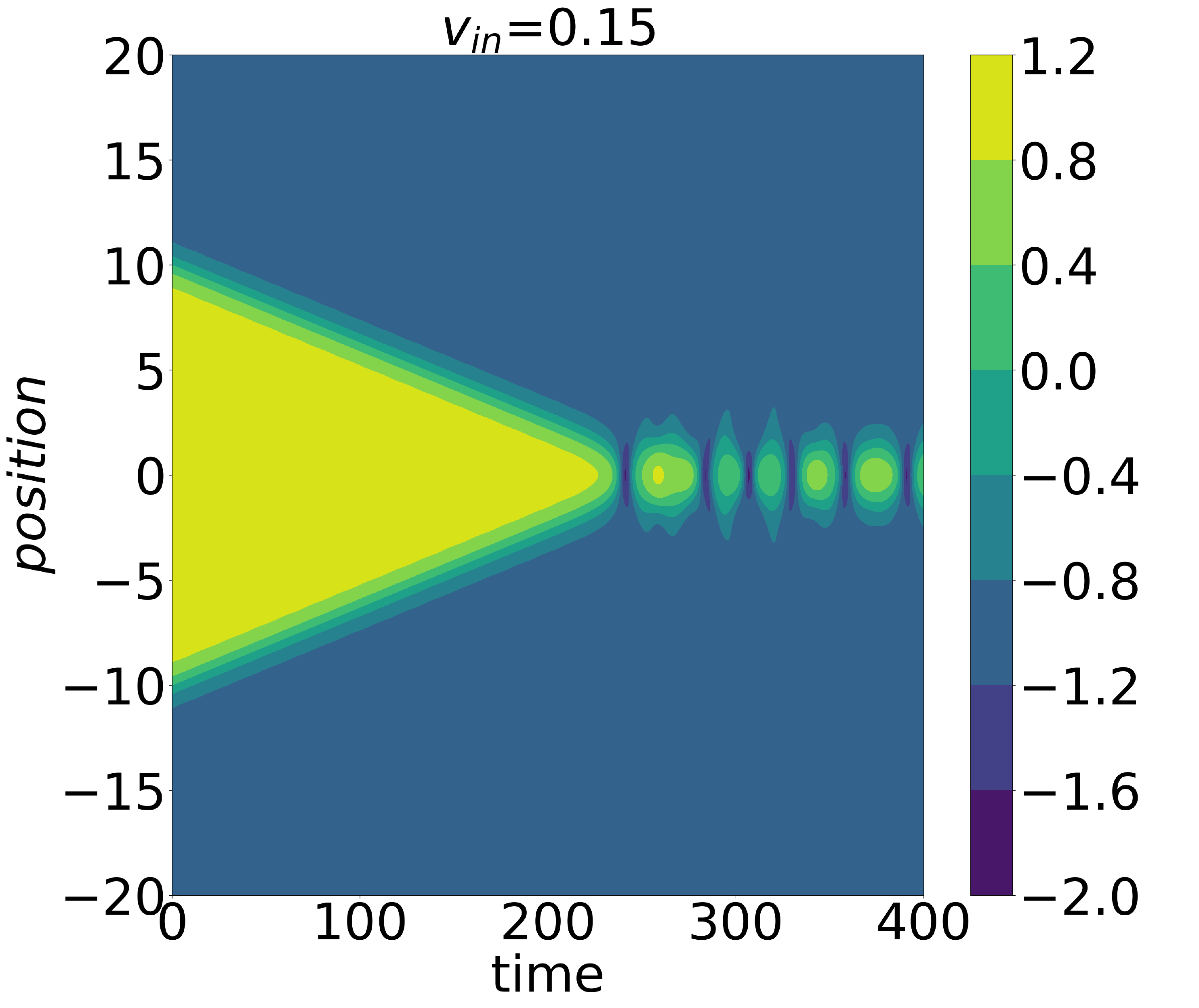} %
\includegraphics[height=4cm,width=5cm]{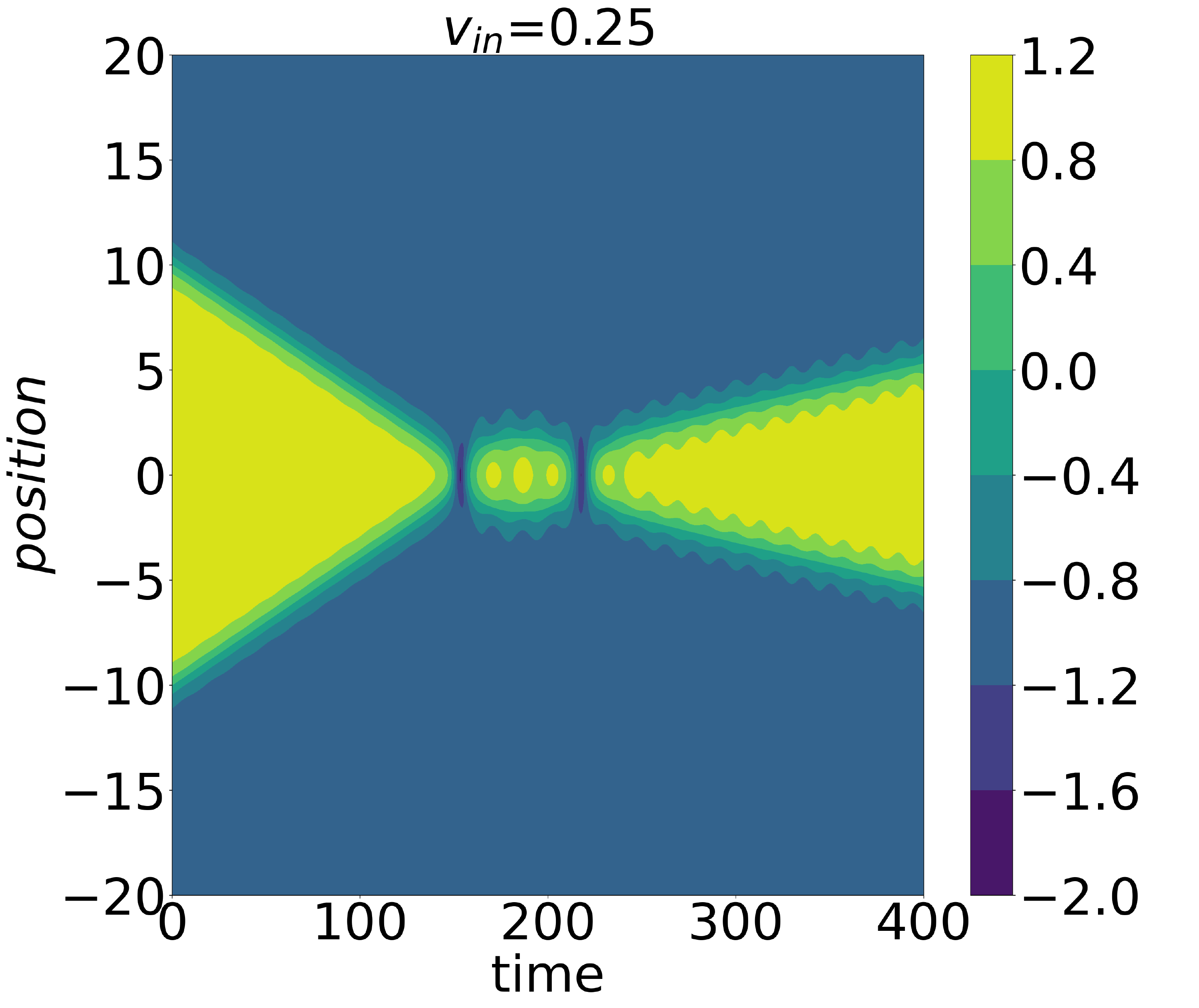}

\vspace{-0.2cm} (a) \vspace{0.3cm}

\includegraphics[height=4cm,width=5cm]{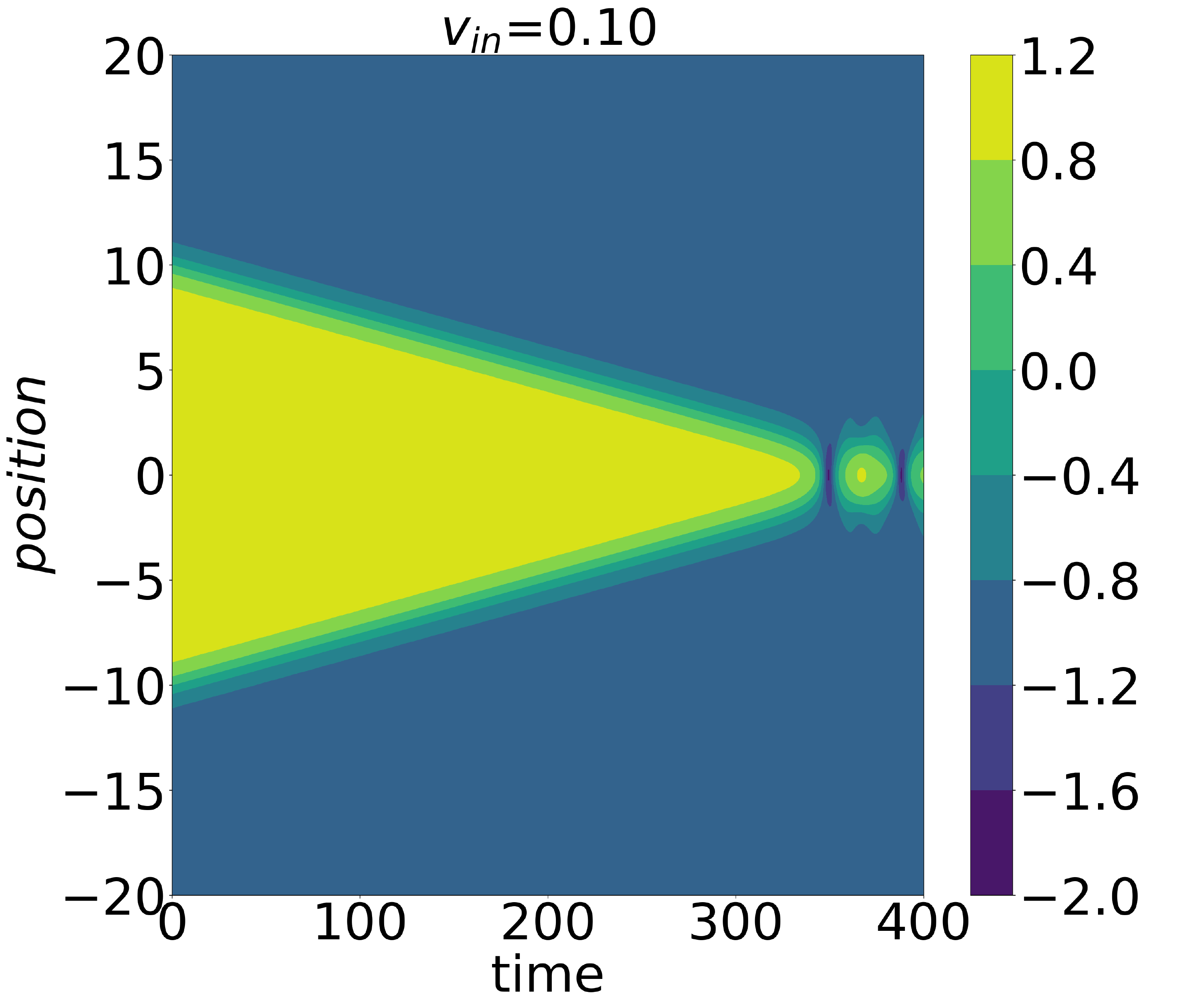} %
\includegraphics[height=4cm,width=5cm]{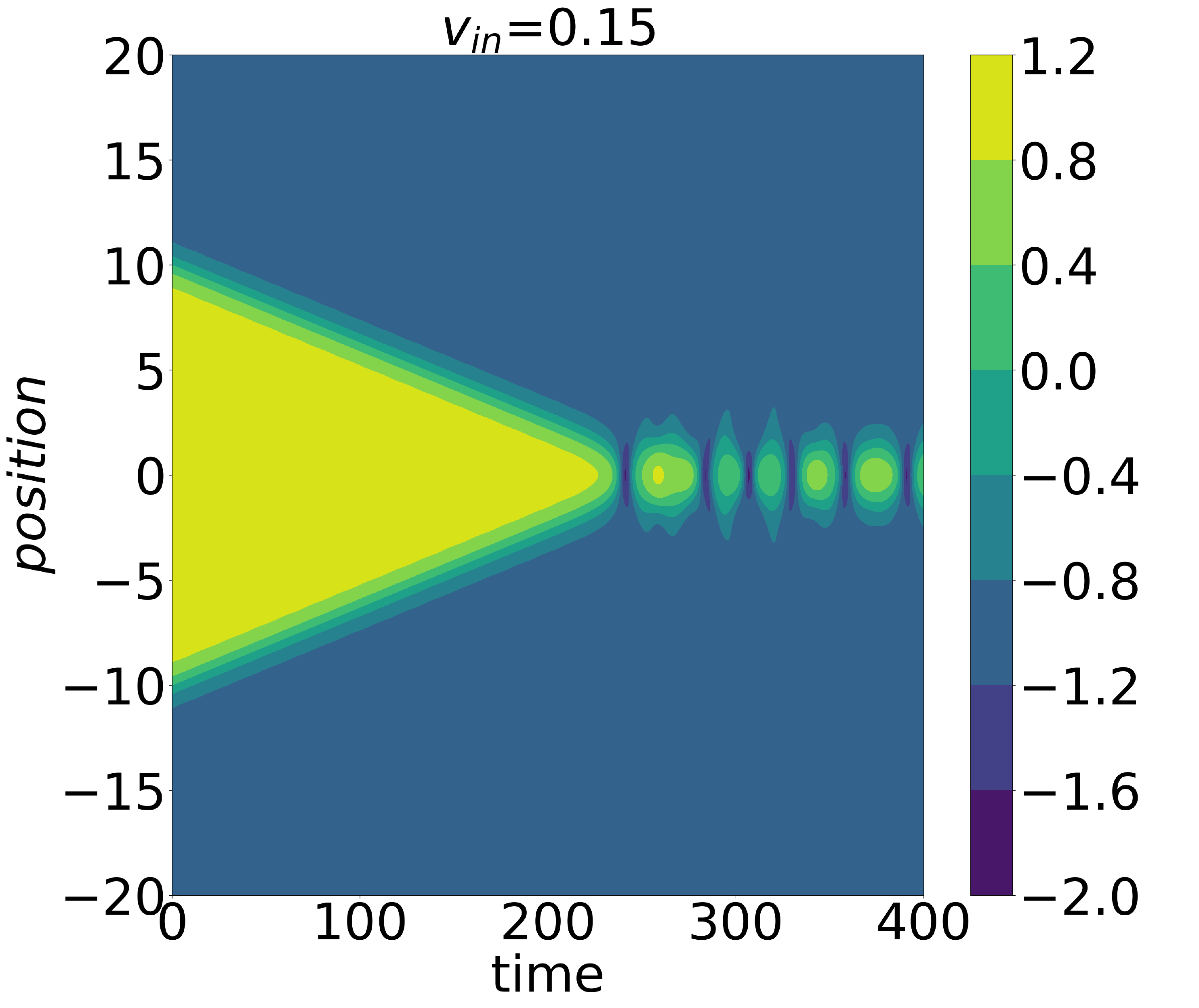} %
\includegraphics[height=4cm,width=5cm]{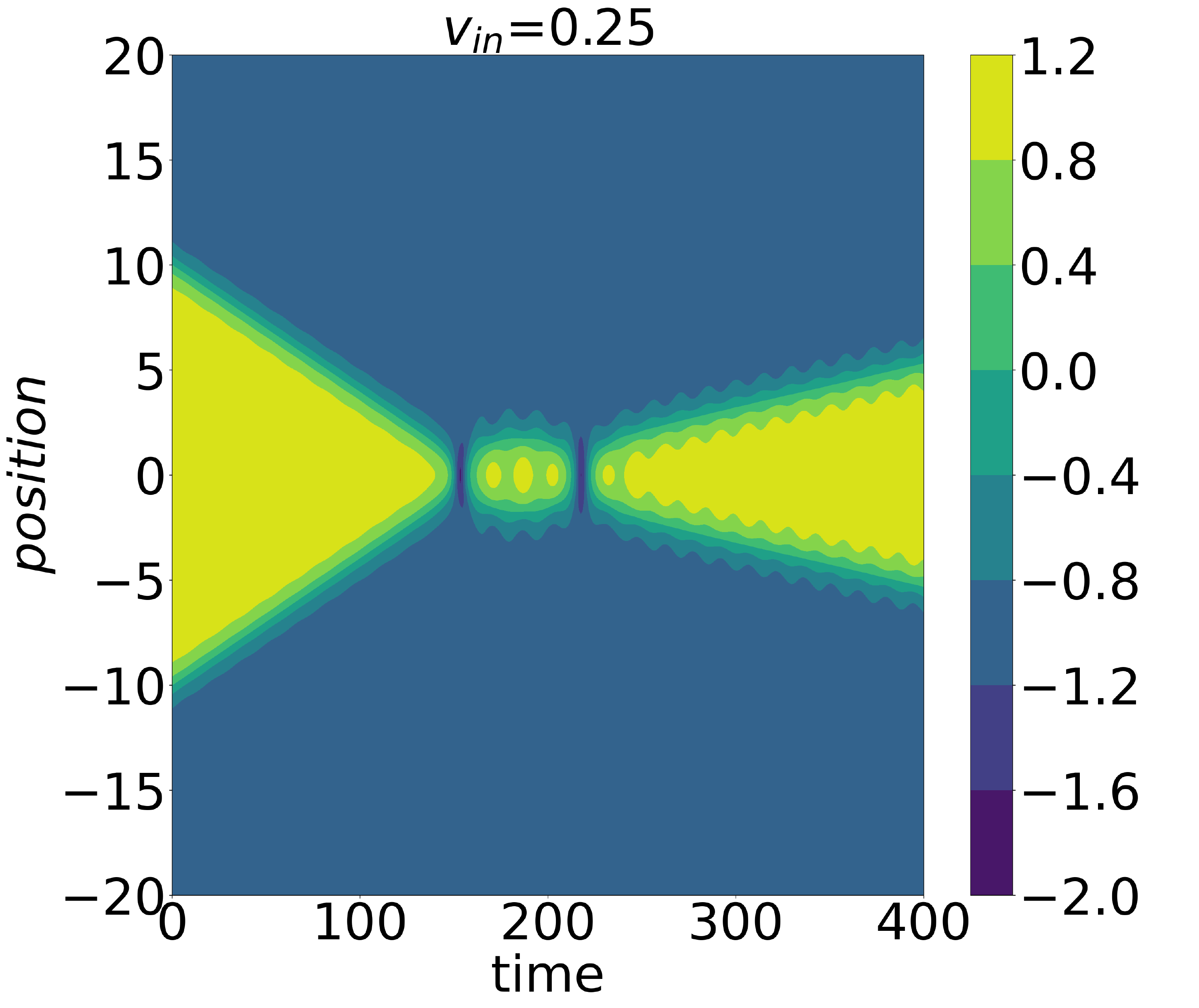}

\vspace{-0.2cm} (b) \vspace{0.3cm}

\includegraphics[height=4cm,width=5cm]{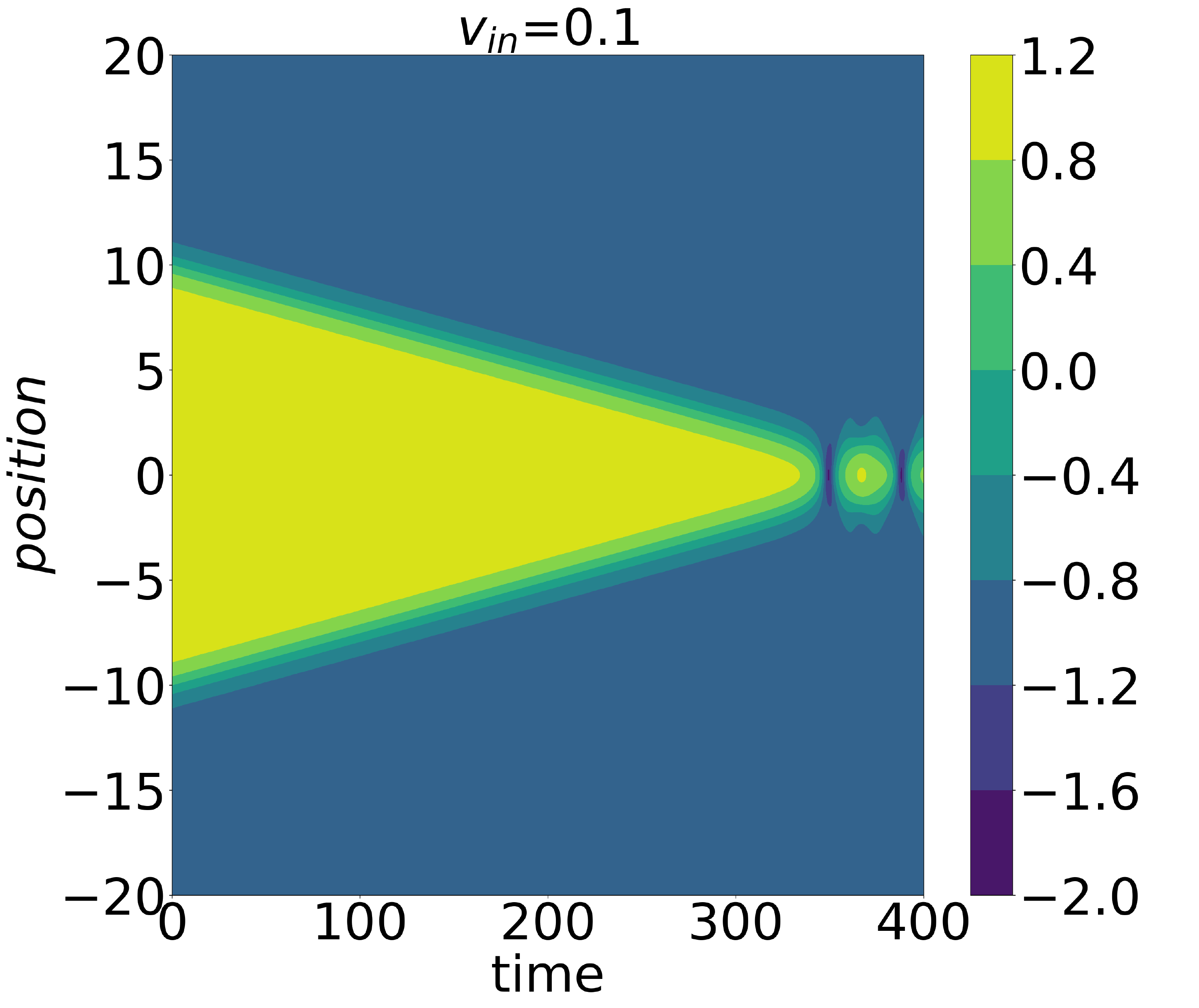} %
\includegraphics[height=4cm,width=5cm]{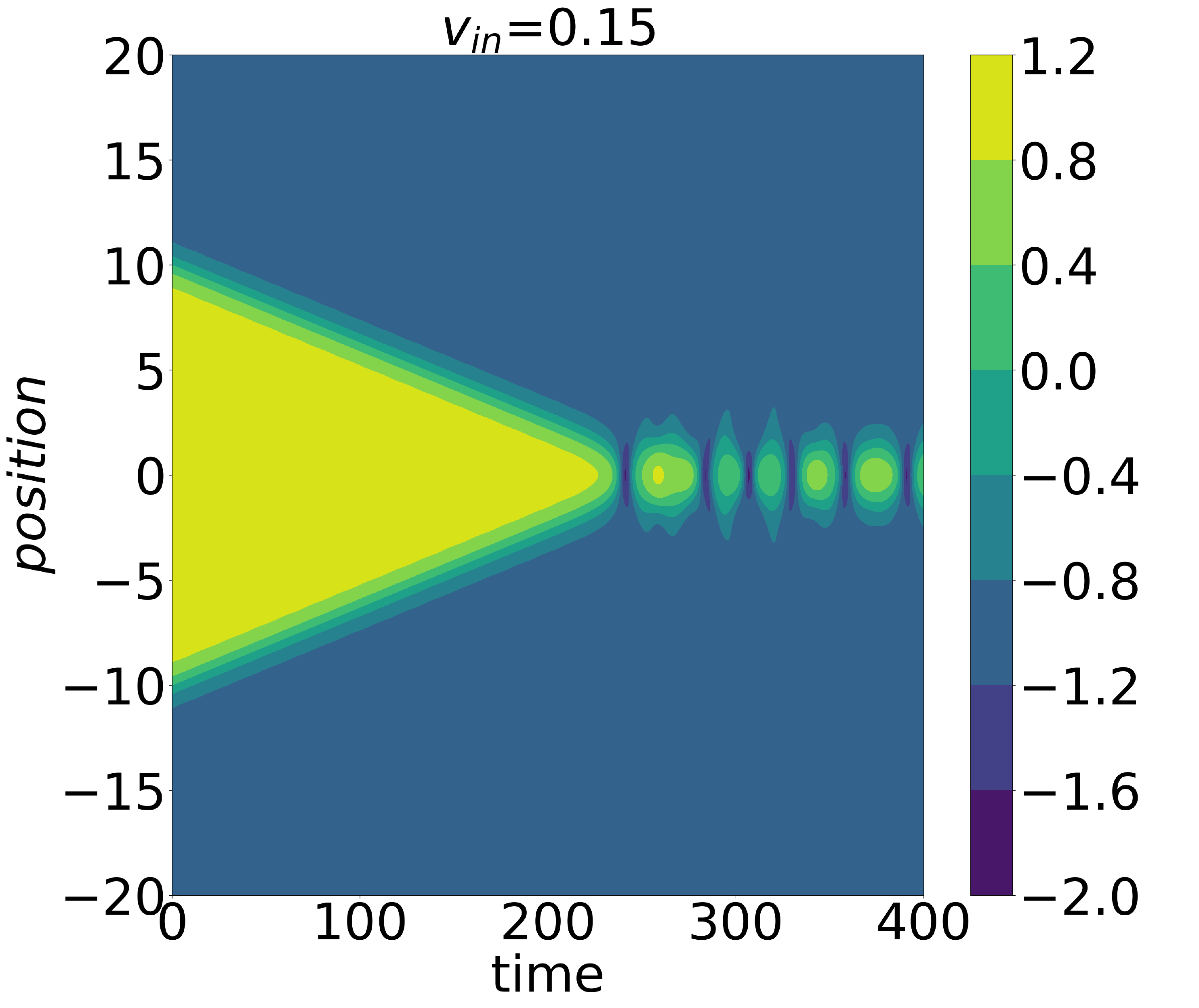} %
\includegraphics[height=4cm,width=5cm]{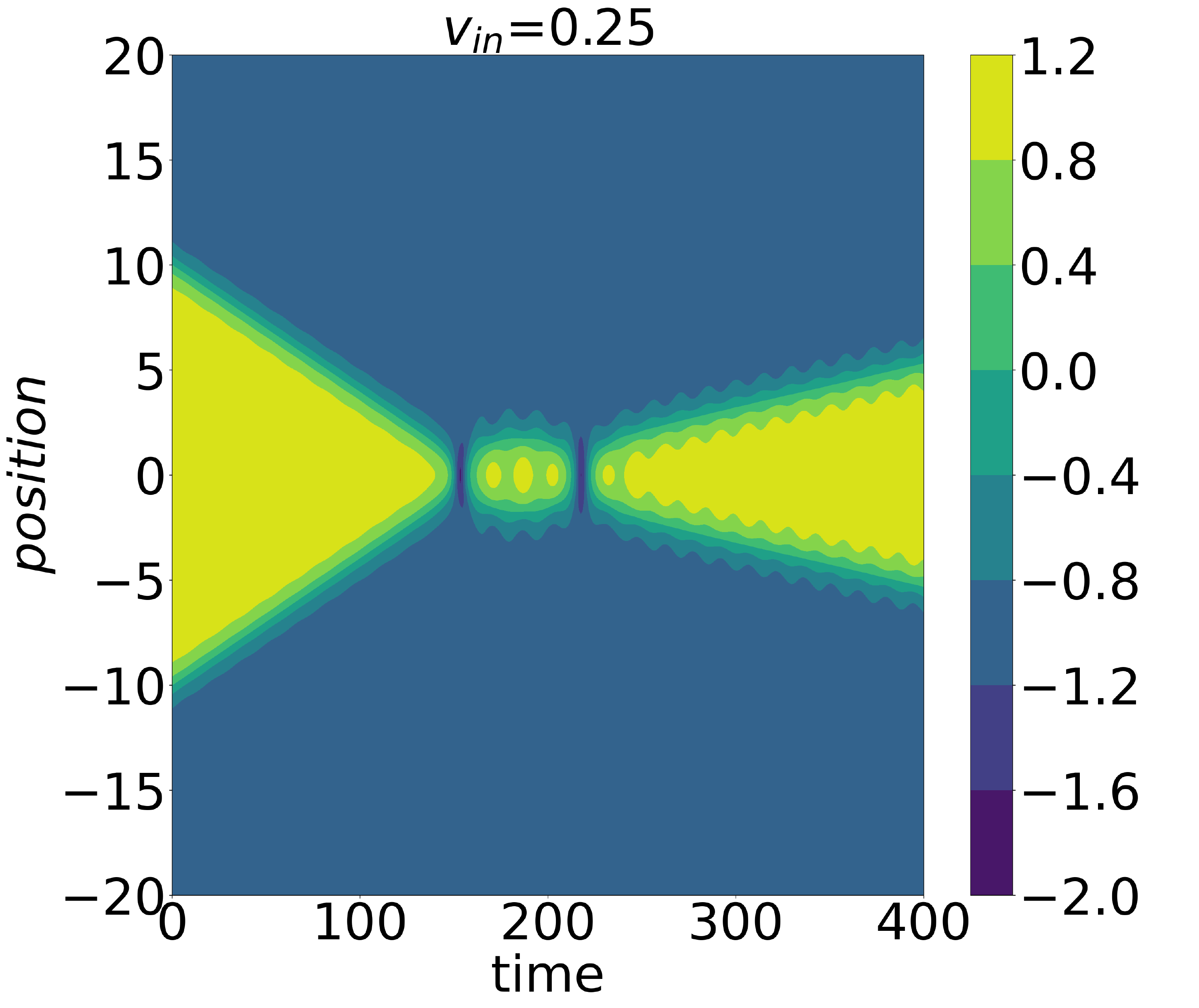}

\vspace{-0.2cm} (c) \vspace{0.3cm}

\includegraphics[height=4cm,width=5cm]{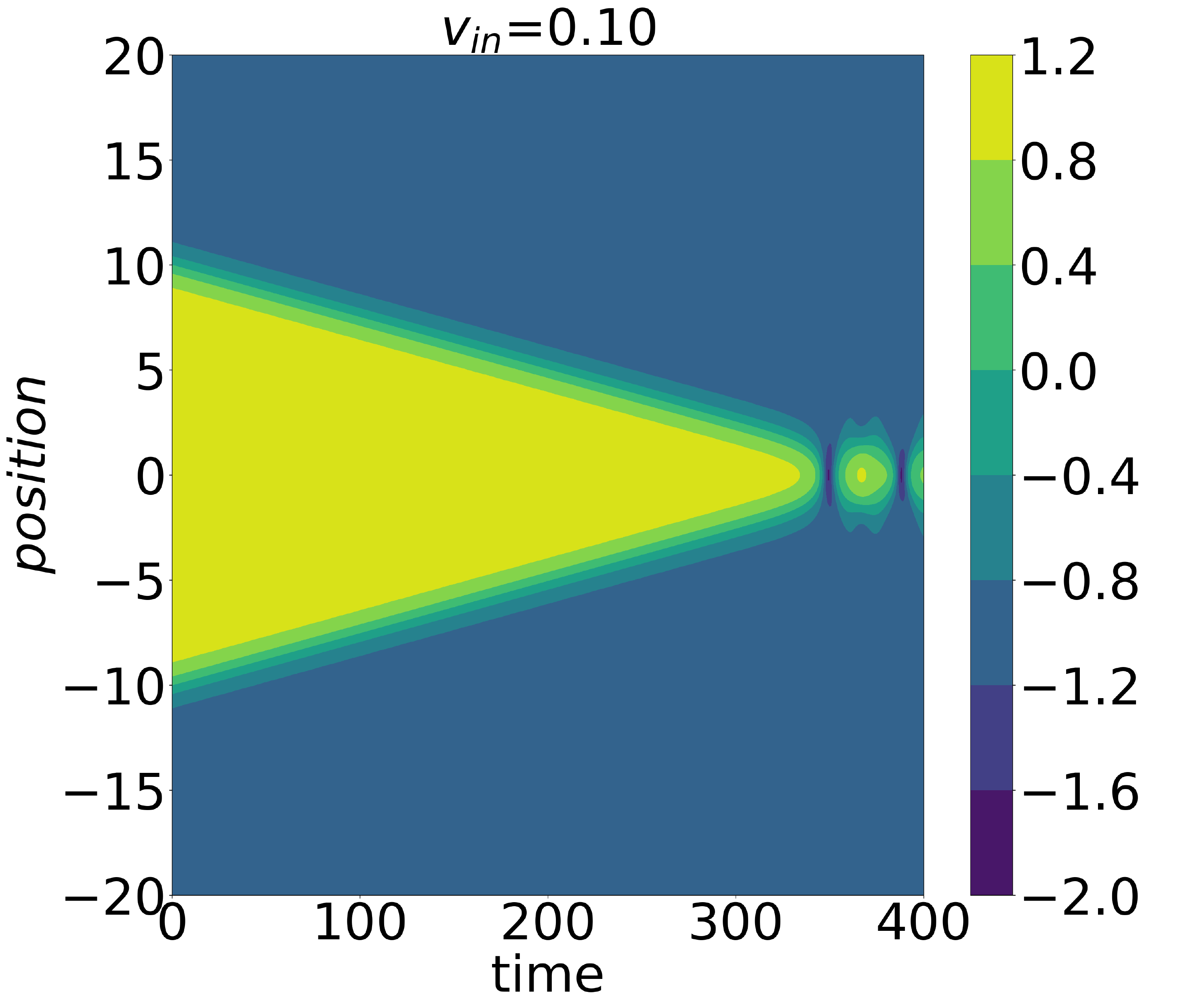} %
\includegraphics[height=4cm,width=5cm]{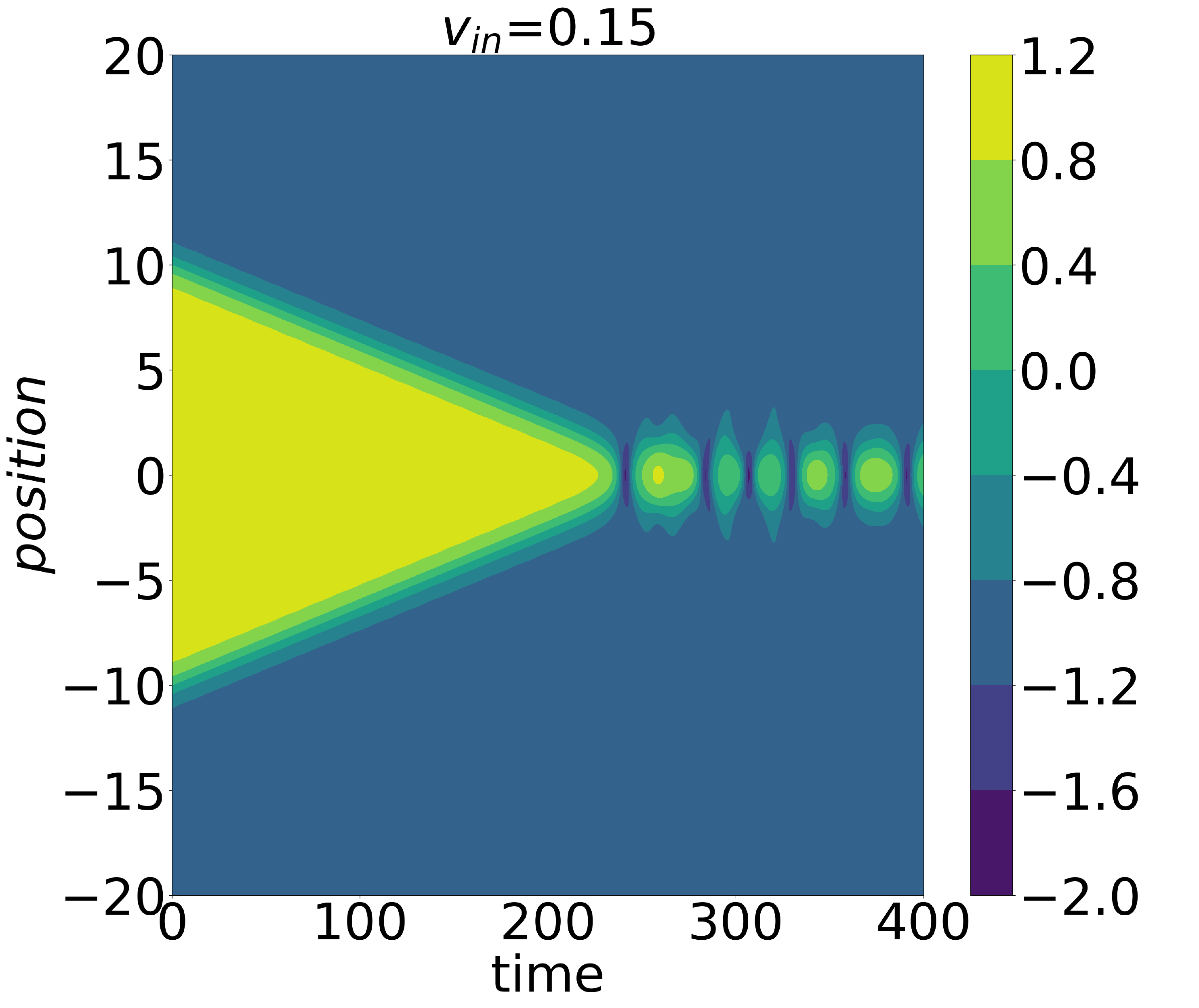} %
\includegraphics[height=4cm,width=5cm]{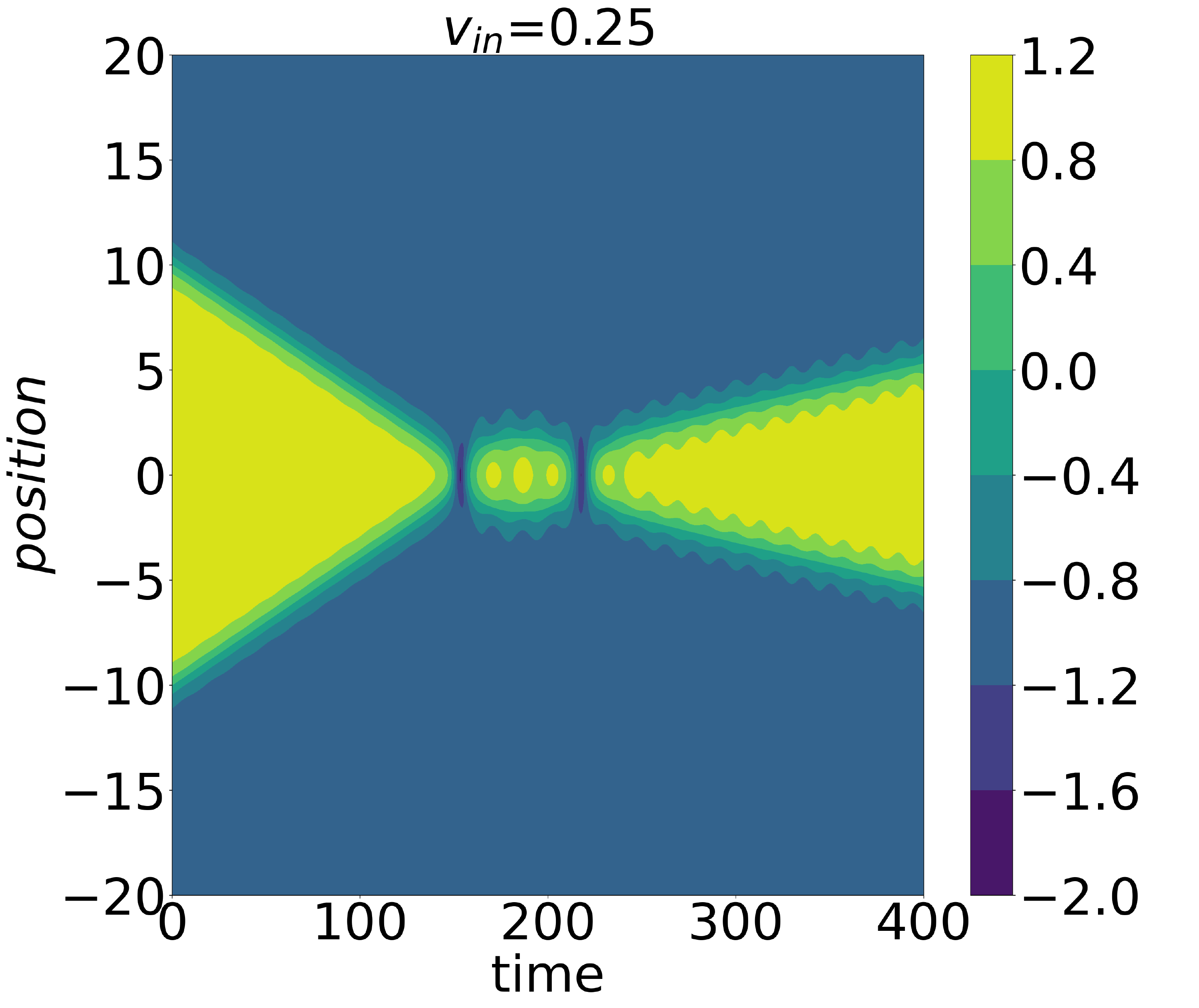}

\vspace{-0.2cm} (d) \vspace{-0.3cm}
\caption{(a) Collision $\lambda=1$ and $m=0.45$. (b)
Collision $\lambda=2$ and $m=0.45$. (c) Collision $\lambda=3$ and $m=0.45$.
(d) Collision $\lambda=4$ and $m=0.45$.} \label{fig10}
\end{figure}

\begin{figure}[!ht]
\centering
\includegraphics[width=8.5cm]{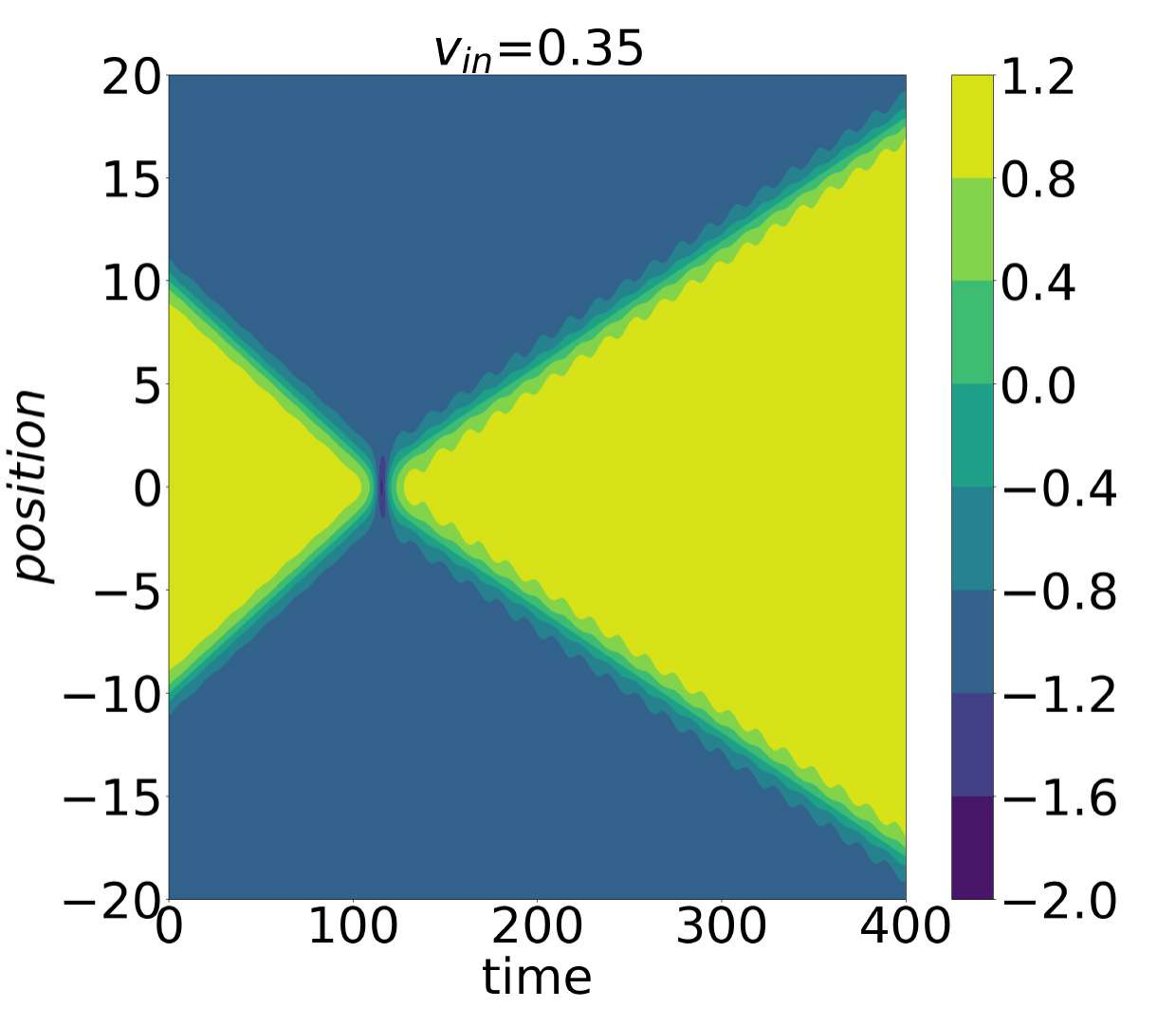} \vspace{-0.3cm}
\caption{Quasi-elastic collision  when $v_{\text{in}}>0.25$ for all $\lambda$.} \label{fig11}
\end{figure}

Figures \ref{fig9} and \ref{fig10} show the analysis of the collisions of the kink-antikink structures for several initial velocity values. One notes that also, throughout the scattering, kink/antikink solutions behave like compacted configurations when $m\to 0.5$ and $\lambda\to\infty$. Besides, it exhibits critical velocities ($v_{\text{cr}}$) that alter the scattering profile. For instance, for $v_{\text{in}}<v_{\text{cr}}^{(1)}\simeq 0.15$, it stays evident that during the temporal evolution, the kink-antikink collide, annihilating each other and radiating energy. Conversely, one obtains an inelastic collision for an initial velocity whose value belongs to the interval $<0.15, 0.25]$. On the other hand, for $v_{\text{in}}>0.25$, the compacton-like configurations, here generated, experiment quasi-elastic collisions; see Fig.  \ref{fig11}.  Thus, the compacton-like configurations in the external scattering process support both inelastic and quasi-elastic collisions, and they manifest as solitary waves with physical characteristics akin to kinks.

By looking closest, the numerical results shown by Figs. \ref{fig9},  \ref{fig10}, and \ref{fig11}, it is noted that for initial velocities smaller than $0.25$, the $K\bar{K}$ pair interacts forming oscillations (bions\footnote{Bions arise during the collision of kinks and have interesting characteristics, such as topological charges, which are a combination of the topological charges of the original kinks. One can interpret the bions as a pair of ``temporary kinks'' that form during a collision.}).For instance, when $v_{\text{in}}=0.10$, the first oscillation emerges. These oscillations increase as the velocity increases, regardless of the values of  $m$ or $\lambda$. However, upon reaching the value of $v_{\text{in}} =0.25$, as time evolution, only three oscillations will occur, and then the $K\bar{K}$ pair scatters with the resonance phenomenon (oscillation amplitudes between $0.8$ and $1.2$) present for a time range greater than $200$. Finally, one notes that when $v_{\text{in}}\geq 0.35$, regardless of the value of $\lambda$, there will be a quasi-elastic scattering process with extreme regions of the $K\bar{K}$ pair undergoing oscillations (resonance) between $0$ and $1.2$

\section{Final remarks \label{sec5}}

The study here has analyzed the existence of $\mathds{Z}_{2}$-kink configurations in a generalized $\phi^4$ model through a noncanonical kinetic term. One uses the generalizing function $f(\phi)=\cos(m\pi\phi)$, which transforms the BPS solutions from a kink configuration (attained for small values of $m$) into a compacton-like profile when $m\rightarrow 0.5$ for all values of $ \lambda$. Indeed, for a fixed $m$, large values of $\lambda$ also engender kink configurations resembling compacton-like solutions. Furthermore, we have observed that the BPS solutions with the same vacuum value $\nu$ and fixed $\lambda$ share the same Bogomol'ny limit (\ref{ENbpsx}), $\mathrm{E}_{\text{BPS}} = 2\sqrt{2} \lambda \nu^3/3$, for all values of $m$ in the $[0,0.5>$ range. Thus, in this sense, the $m$ parameter engenders BPS solutions with infinite degenerescence.

We continue our study by applying the DCC technique to determine the most suitable values for the $m$ parameter of the function $f(\phi)=\cos(m\pi\phi)$ and for the parameter $\lambda$. Thus, through numerical inspection, the DCC leads us to conclude that field configurations with lower complexity are more likely to emerge when $\lambda \simeq 300$ and $m \to 0.5$, i.e., compacton-like structures. It is a significant result because, despite infinite degeneracy, values of $m$ and $\lambda$ exist, describing the best configuration possible for the system.

We conclude our study by analyzing the excitation spectrum of the model for fixed $\nu$ and $\lambda$. Thus, we find a discrete spectrum with translational and vibrational modes. We note that the resonance phenomena are guaranteed to occur during the scattering process of configurations with opposite topological charges. This way, the study reveals that the scattering of the compacton-like solutions presents the following features depending on the value of the initial velocity ($v_{\text{in}}$): for $v_{\text{in}}\simeq 0.15$ they collide and annihilate each other by radiating energy. On the other hand, for an initial velocity occurring in the interval $[0.15, 0.25]$, an inelastic collision happens. These results reveal that the compacton-like configurations behave as solitonic waves with physical aspects similar to the ones arising in the kink/antikink collision processes. Furthermore, the compacton-like solutions collide quasi-elastically when the initial velocity is $v_{\text{in}}>0.25$.

A future perspective naturally involves studies on topological configurations emerging in generalized models with multiple interacting fields. In this context, for instance, we will seek to study structures called vortices in Lorentz-violating scenarios and the presence of nonlinear electrodynamics.

\section{Acknowledgment}

The authors express their gratitude to FAPEMA, CNPq, and CAPES (Brazilian research agencies) for their invaluable financial support. F. C. E. L. is supported by FAPEMA BPD-05892/23. C. A. S. A. is supported by CNPq 309553/2021-0, CNPq/Produtividade. Furthermore, C. A. S. A. is grateful to the Print-UFC CAPES program, project number 88887.837980/2023-00, and acknowledges the Department of Physics and Astronomy at Tufts University for hospitality. R. C. acknowledges the support from the grants CNPq/312155/20 23-9, FAPEMA/UNIVERSAL-00812/19, and FAPEMA/APP-12299/22. F. C. E. L. extends gratitude to F. I. A. Nascimento for the fruitful computational discussion, which proved essential for reaching the project's purpose.

\section{Conflicts of Interest/Competing Interest}

All the authors declared that there is no conflict of interest in this manuscript.



\end{document}